\documentclass[aps,prb,superscriptaddress,amsmath,amssymb,twocolumn,floatfix,reprint]{revtex4-2}
\usepackage{braket}
\usepackage{subfigure}
\usepackage{color}
\usepackage{verbatim}
\usepackage{amsthm,amsmath,amssymb}
\usepackage{mathrsfs}
\usepackage{graphicx}
\usepackage{esint}
\usepackage{mathrsfs}
\usepackage{bm}
\usepackage{hyperref}
\usepackage{mathrsfs}
\usepackage{times}
\usepackage{soul}
\usepackage[normalem]{ulem} 

\newenvironment{localgraphicspath}[1]{
  \graphicspath{#1}
}{}

\definecolor{OliveGreen}{cmyk}{0.64, 0, 0.95, 0.40}
\definecolor{purple}{rgb}{0.6,0,0.5}

\begin{document}

\title{Thermodynamic-Limit Evidence for Chiral Superconductivity Induced by Doping Chiral Topological Phases}

\author{Sen Niu}\email{sen.niu@csun.edu}
\affiliation{Department of Physics and Astronomy, California State University Northridge, California 91330, USA}
\affiliation{School of Physics, Beihang University, Beijing 100191, China}

\author{D. N. Sheng}\email{donna.sheng@csun.edu}
\affiliation{Department of Physics and Astronomy, California State University Northridge, California 91330, USA}

\author{Yang Peng}\email{yang.peng@csun.edu}
\affiliation{Department of Physics and Astronomy, California State University Northridge, California 91330, USA}
\affiliation{Institute of Quantum Information and Matter and Department of Physics, California Institute of Technology, Pasadena, CA 91125, USA}

\begin{abstract}

The emergence of superconductivity from doping strongly correlated chiral topological phases in purely repulsive two-dimensional fermionic systems is a problem of broad and fundamental interest. However, existing numerical evidence has been limited to finite-size studies, and direct thermodynamic-limit evidence for superconducting long-range order has remained lacking. Here we provide such evidence for chiral superconductivity in the triangular Hofstadter-Hubbard model by advancing a simplex tensor-network approach that simultaneously captures superconducting long-range order and chiral topological order in the presence of intrinsic charge fluctuations, a capability that has remained challenging for previous two-dimensional approaches.
We show that a broad intermediate-$U$ chiral spin liquid is separated from the weak-$U$ Chern insulator by a Mott transition, together forming undoped parent chiral topological states. 
Upon hole doping, we identify a uniform chiral superconducting state in the infinite system, characterized by a finite complex pairing order parameter. 
The pairing field exhibits an almost universal phase winding over a broad interaction-doping regime, with a distinct pocket of opposite winding near the Mott criticality. In addition, the entanglement spectrum retains the chiral structure of the parent topological phases while developing additional low-energy branches upon doping.
These results establish that chiral superconductivity emerges robustly from doped chiral topological phases.

\end{abstract}

\date{\today }

\maketitle

{\bf \emph{Introduction.}}~Establishing superconducting (SC) long-range order in purely repulsive two-dimensional fermionic systems remains a central task~\cite{lee2006doping}. 
One appealing scenario envisions superconductivity emerging from chiral topological phases, as proposed in early ideas of anyon superconductivity~\cite{laughlin1988superconducting,laughlin1988relationship,fetter1989random,lee1989anyon,chen1989anyon,wilczek1990fractional,wen1990compressibility} in fractional quantum Hall systems and later extended to lattice settings such as the Kalmeyer-Laughlin chiral spin liquid (CSL)~\cite{kalmeyer1987equivalence,kalmeyer1989theory,wen1989chiral,wen1989effective}. 
Despite extensive theoretical proposals, experimentally relevant CSL realizations remain scarce~\cite{schroeter2007spin,nielsen2012laughlin,he2014chiral,gong2014emergent,bauer2014chiral,liu2016chiral}, limiting concrete microscopic platforms for exploring their doped superconducting descendants.

This long-standing motivation has recently gained renewed relevance in moir\'e-engineered lattice systems~\cite{han2025signatures,xu2025signatures,shi2025doping,pichler2025microscopic}. 
Several microscopic routes toward chiral superconductivity from chiral topological phases have been proposed, including triangular-lattice $t$-$J$-$J_{\chi}$ models~\cite{jiang2020topological,huang2022topological}, spinless Chern-band systems hosting fractional Chern insulators~\cite{guerci2025fractionalization,wang2025chiral}, and Hofstadter–Hubbard models in magnetic flux~\cite{kuhlenkamp2024chiral,divic2025anyon,chen2025topological,kuhlenkamp2025robust}. 
In particular, triangular Hofstadter-Hubbard models offer experimental tunability through large magnetic flux in moir\'e platforms. 
Across these different systems, however, superconductivity has so far been inferred only from finite-size evidence such as pairing correlations or binding energies, and direct thermodynamic-limit evidence based on a complex nonzero order parameter remains absent. 
Establishing spontaneous superconducting order in a genuinely two-dimensional interacting fermionic system therefore constitutes a broader open challenge.

Infinite projected entangled pair states~\cite{verstraete2004renormalization} (iPEPS) and their fermionic generalizations~\cite{barthel2009contraction,kraus2010fermionic,gu2563grassmann,pivzorn2010fermionic,corboz2010simulation,gu2013efficient,ma2024variational} provide an infinite-size tensor-network framework for simulating superconducting and competing orders in two-dimensional $t$–$J$ and Hubbard models~\cite{corboz2011stripes,ctmrg2,poilblanc2014resonating,corboz2016improved,dong2020stable,xu2023competing,yang2023projected,ponsioen2023superconducting,zheng2024competing,zheng2025revealing,zheng2025competing,zhang2025frustration,liu2025accurate}. 
Recent studies have demonstrated that iPEPS can capture chiral topological order in spin models~\cite{hasik2022simulating,budaraju2024simulating,chen2018non,chen20203,chen2021abelian,wang2022emergent,niu2022chiral,xu2023phase,tan20241,puente2025efficient,chen2025simulating,weerda2024fractional,dong2025efficient}, where charge degrees of freedom are absent. 
Addressing the challenge outlined above requires simultaneously treating spontaneous superconducting symmetry breaking and chiral topological order in interacting fermionic systems with intrinsic charge fluctuations, such as those near a Mott transition or at finite doping. 
While iPEPS provides a natural infinite-size framework for this task, its application to such interacting fermionic regimes remains largely unexplored.

\begin{figure}[hbt!]
\begin{localgraphicspath}{{figs/draft_fig/}}
\includegraphics[width=1\columnwidth]{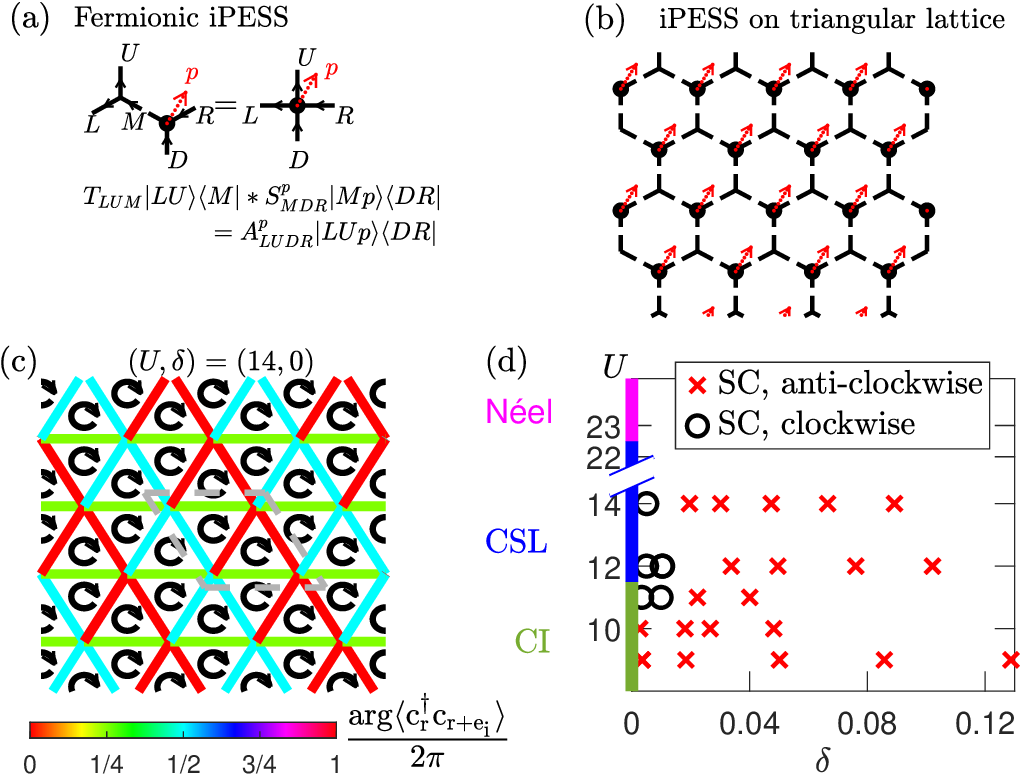}
\caption{{\bf Fermionic iPESS ansatz and phase diagram.} 
(a) Structure of fermionic simplex tensors on a single site. 
(b) Fermionic iPESS representation on the infinite lattice. 
(c) Optimized fermionic iPESS at $D=8$, illustrating the complex Hofstadter hopping pattern and scalar chirality: bond thickness and color encode hopping amplitudes and phases, sizes of black circular arrows indicate scalar chirality, and grey dashed boxes mark the $2\times2$ unit cell. 
(d) Phase diagrams for the insulating ($\delta=0$) and hole-doped ($\delta>0$) regimes.}
\label{fig:sketch}
\end{localgraphicspath}
\end{figure}

In this work, we address this broader issue by studying the $\pi/2$-flux triangular Hofstadter-Hubbard model as a concrete interacting fermionic platform and provide direct thermodynamic-limit evidence for chiral superconductivity. 
To this end, we advance a fermionic infinite projected entangled simplex state (iPESS) approach~\cite{xie2014tensor,li2022magnetization} [Fig.~\ref{fig:sketch}(a-c)], 
a low-rank extension of iPEPS that enables simultaneous resolution of chiral topological order and superconducting long-range order in infinite-size two-dimensional fermionic systems.
We show that the intermediate-$U$ regime hosts a robust CSL, forming an undoped parent phase between the Chern insulator and the $120^{\circ}$ N\'eel state. 
Upon hole doping, the infinite-system variational ground state develops a uniform superconducting state characterized by a finite complex pairing order parameter, thereby establishing superconducting long-range order beyond finite-geometry diagnostics.
Beyond establishing superconducting long-range order in the thermodynamic limit, we further analyze the structure of the doped superconducting state. In particular, the superconducting state exhibits competing pairing patterns characterized by distinct phase winding near the Mott criticality [Fig.~\ref{fig:sketch}(d)]. Finally, we examine the entanglement spectrum of the doped superconducting states across the Mott transition.

{\bf \emph{Model and fermionic iPESS Method.}}~We consider the spin-$1/2$ grand canonical Hofstadter-Hubbard Hamiltonian 
\begin{align}
H&=\sum_{\langle i,j\rangle,\sigma} (t_{ij}c_{i\sigma}^{\dagger} c_{j\sigma}+\rm{h.c.}) \notag\\
&+U\sum_{i}(n_{i\uparrow}-\frac{1}{2})(n_{i\downarrow}-\frac{1}{2}) -\mu\sum_{i,\sigma}n_{i\sigma}
\label{eq:H} 
\end{align}
on the triangular lattice with primitive translation vectors $\boldsymbol{e}_1=(1,0)$, $\boldsymbol{e}_2=(-1/2,\sqrt{3}/2)$, $\boldsymbol{e}_3=\boldsymbol{e}_1+\boldsymbol{e}_2$. Each triangle is pierced by $\pi/2$ magnetic flux, and we adopt the gauge $t_{\boldsymbol{r},\boldsymbol{r}+\boldsymbol{e}_1}=it$, $t_{\boldsymbol{r},\boldsymbol{r}+\boldsymbol{e}_2}=(-1)^{x}t$, $t_{\boldsymbol{r},\boldsymbol{r}+\boldsymbol{e}_3}=(-1)^{x-1}t$ from Ref.~\cite{kuhlenkamp2024chiral}, with $\boldsymbol{r}=(x\boldsymbol{e}_1,y\boldsymbol{e}_2)$, $t=1$. In the non-interacting limit, the model has a $2\times 1$  unit cell and exhibits a pair of particle-hole symmetric $C=\pm1$ Chern bands. The average filling fraction $\nu=1-\delta$ is tuned by chemical potential $\mu$, with $\mu=0$ giving half-filling $\nu=1$ due to  particle-hole symmetry.

To tackle the fermionic model on the infinite triangular lattice, we introduce the fermionic iPESS ansatz, which combines the simplex structure~\cite{schuch2012resonating,xie2014tensor,li2022magnetization} with fermionic tensor techniques~\cite{barthel2009contraction,kraus2010fermionic,gu2563grassmann,pivzorn2010fermionic,corboz2010simulation,gu2013efficient}.  We define a rank-3 virtual simplex $\hat{T}=T_{LUM}|LU\rangle\langle M|$ tensor at the center of each up-triangle and a rank-4 physical $\hat{S}=S_{MDR}^{p}|Mp\rangle\langle DR|$ tensor on each lattice site, as shown in Fig.~\ref{fig:sketch} (a)-(b). Here $L,D,R,U,M$ denote virtual fermionic modes with bond dimension $D$, and $p$ indexes the four-dimensional local Hubbard Hilbert space. Imposing $Z_2$ fermion parity symmetry in both tensors and Hamiltonian operators avoids non-local Jordan-Wigner strings on the infinite 2D lattice. Contracting the $M$ index between neighboring $\hat{T}$ and $\hat{S}$ tensors shows that iPESS is a subclass of standard iPEPS, 
with the simplex structure treating entanglement among three sites within a triangle equally and better preserving point group symmetries.

Because the fermionic Hubbard model has an enlarged local Hilbert space, accessing sufficiently large bond dimensions is crucial. The rank-3 simplex structure of iPESS substantially reduces memory cost compared to conventional rank-4 iPEPS tensors, enabling large-scale variational simulations at larger $D$ and larger unit-cells. To further enhance efficiency while preserving spin symmetry, we impose spin-$\rm{SU}(2)$ symmetry on the tensors (except in the large-$U$ magnetic regime) without fixing particle number in the grand canonical ensemble. The tensors are optimized variationally~\cite{liao2019differentiable}, and expectation values are evaluated using the corner-transfer-matrix renormalization group (CTMRG)~\cite{ctmrg1,ctmrg2}, which incorporates long-range correlations essential for accurately capturing chiral topological phases. Additional algorithmic details and benchmark analyses are provided in the Supplemental Material (SM)~\cite{SM}.

\begin{figure}[hbt!]
\begin{localgraphicspath}{{figs/draft_fig/}}
\includegraphics[width=1\columnwidth]{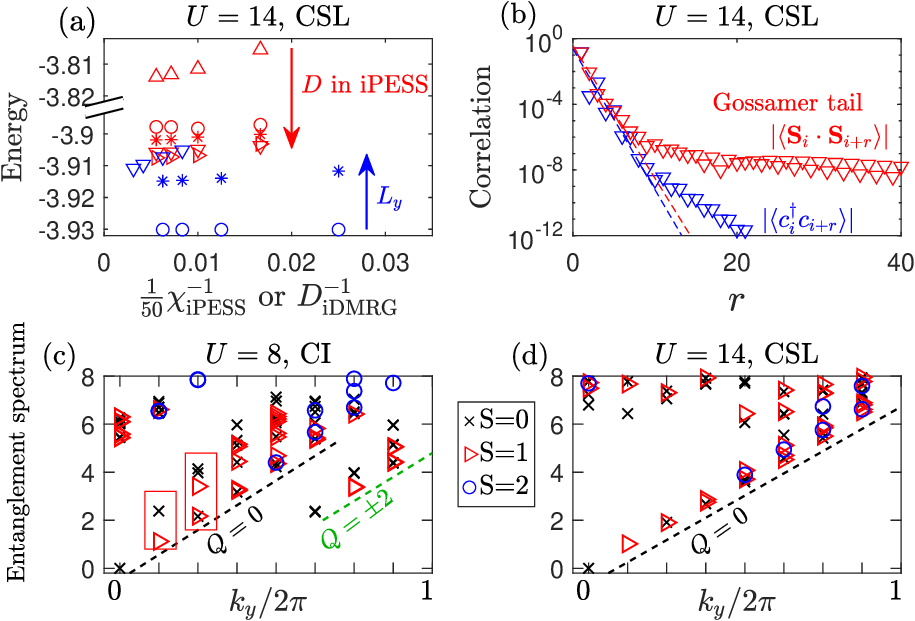}
\caption{{\bf Intermediate-$U$ chiral spin liquid (CSL) phase.} 
(a) Energy convergence of finite-$D$ iPESS ($D=6,8,10,12,14$) compared with finite-$L_y$ iDMRG ($L_y=4,6,8$). $\chi_{\rm{iPESS}}$ denotes the CTMRG environment dimension controlling contraction accuracy. 
(b) Correlation functions obtained from iPESS for $(D,\chi_{\rm{iPESS}})=(12,160)$, including long-distance gossamer tails in the spin-spin channel. 
(c)-(d) Entanglement spectra on a cylinder of width $L_y=8$ for $(D,\chi_{\rm{iPESS}})=(12,60)$ with antiperiodic boundary condition.}
\label{fig:compare_CI_CSL}
\end{localgraphicspath}
\end{figure}

{\bf \emph{Undoped chiral phases and robustness of the chiral spin liquid.}}~At half filling, the model hosts a CSL in an intermediate-$U$ regime~\cite{kuhlenkamp2024chiral}, serving as the undoped parent state for the SC phase discussed below. In the weak-$U$ limit the system reduces to a gapped Chern insulator, separated from the CSL by a Mott transition. 
To assess the reliability of the fermionic iPESS ansatz for such strongly correlated chiral states, we examine the convergence of the $2\times2$ variational solution. The finite-$D$ energies for $U=14$ are shown in Fig.~\ref{fig:compare_CI_CSL}(a). The iPESS energy decreases systematically with increasing $D$, while iDMRG results approach the thermodynamic limit from below with increasing cylinder width $L_y$, demonstrating consistent convergence between the two methods.
The optimized tensors reproduce the complex hopping structure of Eq.~\eqref{eq:H} and exhibit nearly uniform hopping amplitudes and finite scalar chirality $\langle\chi_{ijk}\rangle$, confirming that iPESS preserves lattice symmetry. As expected for chiral topological states at finite $D$, spin-spin correlation functions display characteristic gossamer tails~\cite{hasik2022simulating,niu2024chiral} that decay slower than any exponential form [Fig.~\ref{fig:compare_CI_CSL}(b)], consistent with the presence of topological spin degrees of freedom in a Mott insulating state.

Topological properties of chiral states are reflected in the entanglement spectrum (ES) via the Li-Haldane conjecture~\cite{li2008entanglement}. 
We compute the ES by placing the optimized iPESS with $2\times2$ unit cells on finite-width cylinders~\cite{cirac2011entanglement,poilblanc2012topological,poilblanc20162}, see detailed fermionic algorithm in SM~\cite{SM}. 
The ES on a width-$8$ cylinder are shown in Fig.~\ref{fig:compare_CI_CSL}(c)-(d). 
For $U=8$, charge fluctuations are not strongly suppressed, as seen by the dominant $Q=0$ branch and the degenerate $Q=\pm2$ branches marked by charge $Q$. Low-lying levels of a single branch  match the spinful CI pattern 
$(0), (0)\oplus (1)$, $3\times(0)\oplus 2\times(1),\ldots$, where $(S)$ denotes the SU(2) spin-$S$ representation and the prefactor indicates its multiplicity. 
For $U=14$ and large enough bond dimensions $D\ge 8$, a single low-lying branch identifies the Kalmeyer-Laughlin CSL, 
with reduced level counting $(0), (1), (0)\oplus (1), (0)\oplus 2\times(1),\ldots$ agreeing with $\rm{SU}(2)_1$ conformal field theory~\cite{gawedzki1990}. 
Together, the ES and these finite-$D$-induced gossamer tails provide a key diagnostic that fermionic iPESS faithfully captures the topology with the presence of charge fluctuation.

\begin{figure}[hbt!]
\begin{localgraphicspath}{{figs/draft_fig/}}
\includegraphics[width=1\columnwidth]{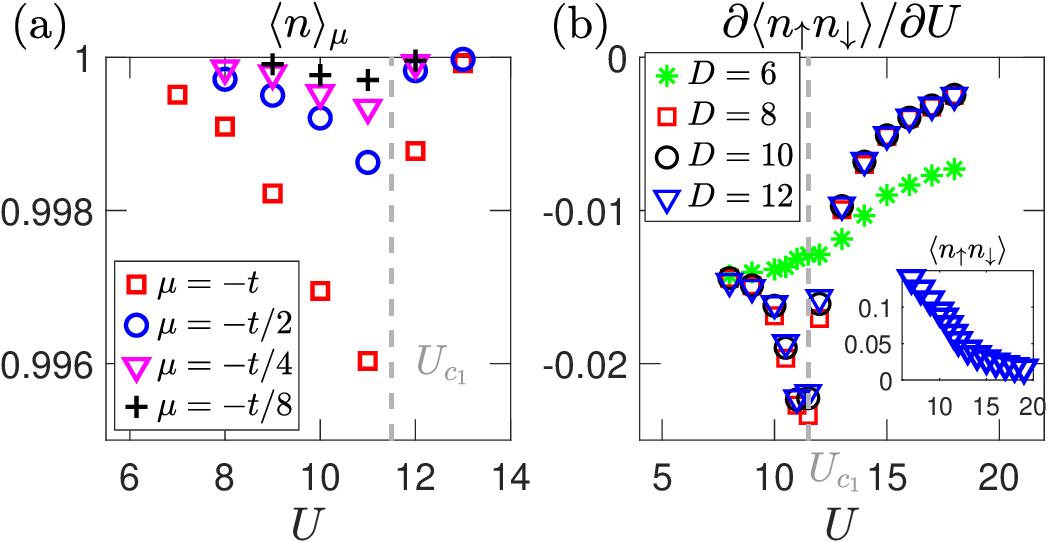}
\caption{{\bf Mott transition between the Chern insulator and chiral spin liquid.} 
The critical interaction $U_{c_1}$ is indicated by the grey dashed line. 
(a) Filling $\langle n \rangle$ versus chemical potential $\mu$ near half filling, showing enhanced charge response across the transition. 
(b) Double occupancy $\langle n_{i\uparrow} n_{i\downarrow} \rangle$ (inset) and its derivative with respect to $U$ at half filling.}
\label{fig:Mott}
\end{localgraphicspath}
\end{figure}

We further examine the stability of the undoped parent CSL. We first determine the critical $U_{c_1}$ between CI and CSL. The sensitivity of the mean occupation $\bar{n}=\langle n \rangle_{\mu}$ to chemical potential $\mu$ is shown by Fig.~\ref{fig:Mott}(a). 
Ref.~\cite{divic2025anyon} proposes a transition between band and Mott insulators, where the charge-$2e$ gap closes while the spin gap remains open.
The occupation $\bar{n}$ exhibits a minimum near $U_{c_1}\sim 11.5$, rapidly approaching $1$ as $\mu\rightarrow 0$ on both sides. We identify $U_{c_1}$ as the 
critical point since CI and Mott insulators are incompressible ($\chi_{e}=\partial \bar{n}/\partial\mu=0$) except at charge gap closure.
Double occupancy in Fig.~\ref{fig:Mott}(b), reflecting charge fluctuations via $\langle n_{i\uparrow}n_{i\downarrow} \rangle=\langle (n-\bar{n})^2 \rangle /2$ for $\bar{n}=1$, 
consistently shows a sudden suppression at $U_{c_1}$, with a singularity in its derivative for $D\ge 8$. 
At large $U$, the system undergoes a transition to a $120^{\circ}$ N\'eel ordered phase at $U_{c_2} \approx 22.5$, as indicated by finite-correlation-length scaling of the magnetization (see SM). These results confirm the CSL as a robust undoped parent phase for a wide parameter regime in thermodynamic limit in the Hofstadter-Hubbard model, contrasting with the narrow CSL regime in the standard triangular Hubbard model~\cite{szasz2020chiral,chen2022quantum,tocchio2021hubbard}.

\begin{figure}[hbt!]
\begin{localgraphicspath}{{figs/draft_fig/}}
\includegraphics[width=1\columnwidth]{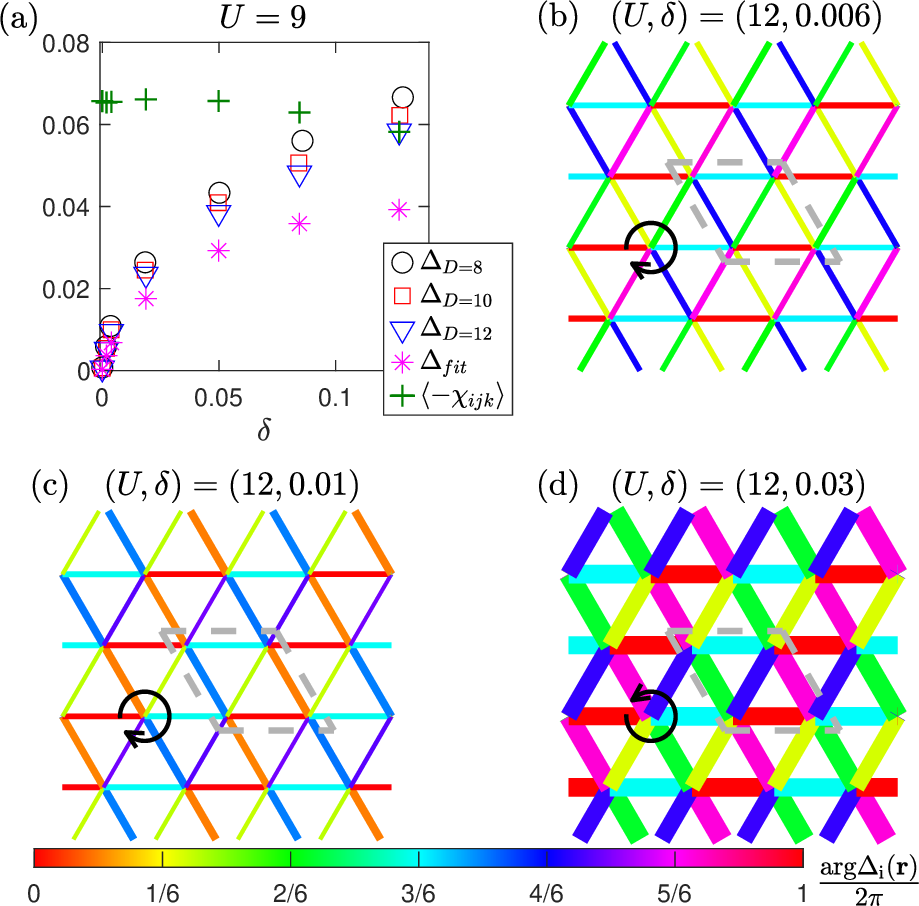}
\caption{{\bf Superconducting order parameter and pairing patterns versus hole doping $\delta$.} 
(a) Average pairing amplitude $\Delta$ and scalar spin chirality as functions of $\delta$ for $U=9$, including $1/D$ extrapolation. 
(b)-(d) Representative pairing configurations in the doped Mott regime at $U=12$ and $D=12$. Bond thickness and color encode the magnitude and phase of $\Delta_i(\mathbf{r})$, respectively. Black circular arrows indicate the phase winding direction. Grey dashed boxes denote the $2\times2$ unit cell.}
\label{fig:SC}
\end{localgraphicspath}
\end{figure}

{\bf \emph{Superconducting order parameter in the thermodynamic limit.}}~We next investigate superconductivity in the lightly hole-doped regime near the Mott transition $U_{c_1}$. Finite-$\delta$ states are efficiently simulated using an $\mathrm{SU}(2)$-symmetric $2\times2$ iPESS ansatz, with optimized half-filled solutions serving as initial states for variational optimization. The superconducting order parameter is defined as $\Delta=\overline{|\Delta_i(\mathbf{r})|}$, where $\Delta_i(\mathbf{r})=\langle c_{\mathbf{r},\uparrow}^{\dagger}c_{\mathbf{r}+\mathbf{e}_i,\downarrow}^{\dagger}-c_{\mathbf{r},\downarrow}^{\dagger}c_{\mathbf{r}+\mathbf{e}_i,\uparrow}^{\dagger}\rangle/\sqrt{2}$ denotes the complex nearest-neighbor singlet pairing amplitude. We obtain a finite pairing amplitude $|\Delta|$ over a broad interaction range spanning both the weak-$U$ Chern-insulating regime and the intermediate-$U$ CSL regime around $U_{c_1}$, which increases with doping for $0<\delta\le0.15$ [regions marked in Fig.~\ref{fig:sketch}(d)]. A representative analysis at $U=9$ is shown in Fig.~\ref{fig:SC}(a): as the bond dimension $D$ increases, $\Delta$ decreases slightly due to enhanced quantum fluctuations, yet remains finite after linear extrapolation in $1/D$, establishing superconducting long-range order in the thermodynamic limit. 
Alternative linear, quadratic, and power-law extrapolations are compared in the Supplemental Material.
To exclude competing symmetry-breaking orders, we further perform unrestricted $Z_2$ iPESS simulations with enlarged unit cells including $4\times4$, $6\times 2$, and $2\times 6$. These yield consistent $\Delta_i(\mathbf{r})$ values while charge-density-wave and magnetic moments remain negligible in this doping range~\cite{SM}, confirming a robust uniform superconducting ground state.

{\bf \emph{Pairing symmetry and winding competition.}}~Having established a nonzero pairing amplitude, we then analyze the pairing symmetry and its evolution across the phase diagram. 
In the optimized iPESS solution, the nearest-neighbour pairing exhibits a staggered structure along the $y$ direction, enlarging the original $2\times1$ unit cell to a $2\times2$ periodicity. To facilitate direct comparison with pair-pair correlations reported in previous finite-size studies~\cite{divic2025anyon,chen2025topological,kuhlenkamp2025robust}, we perform a gauge transformation to a $C_6$-symmetric imaginary gauge with a $2\times2$ unit cell~\cite{SM}. The resulting nearest-neighbour pairing pattern $\Delta_i(\mathbf{r})$ is shown in Fig.~\ref{fig:SC}(b)-(d), where bond thickness and color encode the magnitude and phase of $\Delta_i(\mathbf{r})$, respectively. 
Over a broad region of the $U$-$\delta$ phase diagram, the pairing phases evolve smoothly from $0$ to $2\pi$ around elementary plaquettes, forming a uniform anti-clockwise winding [Fig.~\ref{fig:SC}(d)]. This nearly universal winding pattern characterizes the dominant chiral superconducting state and is consistent with the pair-pair correlations from previous finite-size numerical studies in the canonical ensemble~\cite{divic2025anyon,chen2025topological,kuhlenkamp2025robust}.

On the Mott side of the phase diagram ($U \gtrsim U_{c_1}$), the pairing structure undergoes a qualitative reorganization as the doping is reduced toward half filling. 
An intermediate crossover regime [Fig.~\ref{fig:SC}(c)] displays a non-$C_6$-symmetric distribution of pairing amplitudes and phases, reflecting the coexistence of distinct winding tendencies.
At very low doping $\delta < 0.01$, the system develops a uniform superconducting state with a clockwise winding [Fig.~\ref{fig:SC}(b)], forming a small but distinct pocket near the Mott criticality that was not resolved in previous finite-size studies. 
The evolution of winding patterns over the full $U$-$\delta$ phase diagram is summarized in Fig.~\ref{fig:sketch}(d).
A finer doping scan of the real-space pairing patterns is presented in the Supplemental Material, showing that the winding reversal proceeds through an intermediate regime with nonuniform amplitudes and phase increments.
The emergence of this opposite-winding pocket indicates a competition between distinct pairing patterns on the Mott side, whose microscopic origin and connection to the Mott transition warrant further investigation.
To clarify whether this winding reorganization modifies the global topological properties, we next examine the topological characteristics of the superconducting state.

\begin{figure}[hbt!]
\begin{localgraphicspath}{{figs/draft_fig/}}
\includegraphics[width=1\columnwidth]{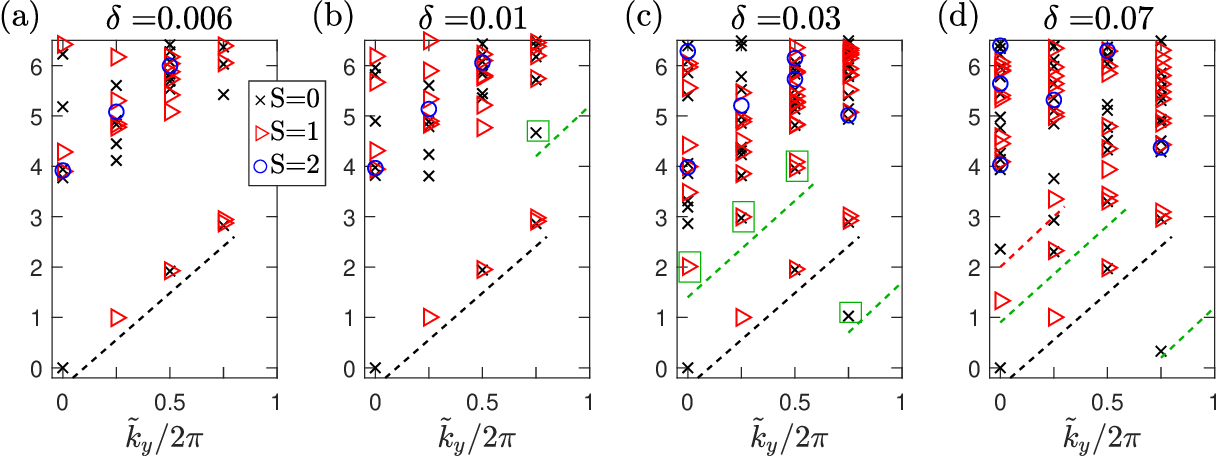}
\caption{{\bf Entanglement spectrum of doped superconducting states at $U=12$ on a cylinder of width $L_y=8$.} The green dashed line indicates an additional low-lying branch that emerges with increasing $\delta$.
Data are shown for $(D,\chi_{\mathrm{iPESS}})=(8,60)$.
$\tilde{k}_y$ is the momentum quantum number defined with respect to the two-site translation symmetry $T_y^2$ along the $y$ direction.
In the $\delta \rightarrow 0$ limit, where superconducting order vanishes and single-site translation symmetry $T_y$ is restored, $\tilde{k}_y = 2k_y$.}
\label{fig:SC_ES}
\end{localgraphicspath}
\end{figure}

{\bf \emph{Topological signatures of doped chiral spin liquid.}}~
Local order parameters alone do not fully determine the global topological character of a superconducting state. We therefore analyze the entanglement spectrum of the doped states. We place the optimized $2\times2$-cell iPESS on cylinders of width $L_y=8$, where the sign structure of the superconducting order parameter reduces the translation symmetry from $T_y$ to $T_y^2$ and folds the Brillouin zone. As shown in Fig.~\ref{fig:SC_ES}(a)-(d) for $U=12$, corresponding to the distinct local pairing patterns in Fig.~\ref{fig:SC}(b)-(d), the ES exhibits a robust chiral structure across different doping rates. 
In particular, the superconducting states retain the same chiral level counting $(0), (1), (0)\oplus (1)$, $(0)\oplus 2\times(1),\ldots$  as the parent chiral spin liquid, characterized by a single unidirectional dispersion without counter-propagating modes. In addition to this conformal tower (black dashed line), an extra low-energy branch (green dashed line) emerges with increasing doping, reflecting enhanced charge fluctuations associated with $\mathrm{U}(1)$ symmetry breaking in the superconducting state.

The persistence of a chiral ES across both winding configurations suggests that the doped superconducting states inherit the chiral topological character of the undoped parent state, consistent with the quantized spin pumping observed in previous finite-size studies~\cite{chen2025topological,kuhlenkamp2025robust}. 
Furthermore, the finite spin-scalar chirality $\chi_{ijk}$ [Fig.~\ref{fig:SC}(a)] and long-range gossamer correlation tails (see SM) provide additional evidence for the robustness of chiral topological characteristics in the doped regime. 
Collectively, these results indicate that the superconducting ground state from doping CSL possesses nontrivial chiral topological characteristics.

\begin{figure}[hbt!]
\begin{localgraphicspath}{{figs/draft_fig/}}
\includegraphics[width=1\columnwidth]{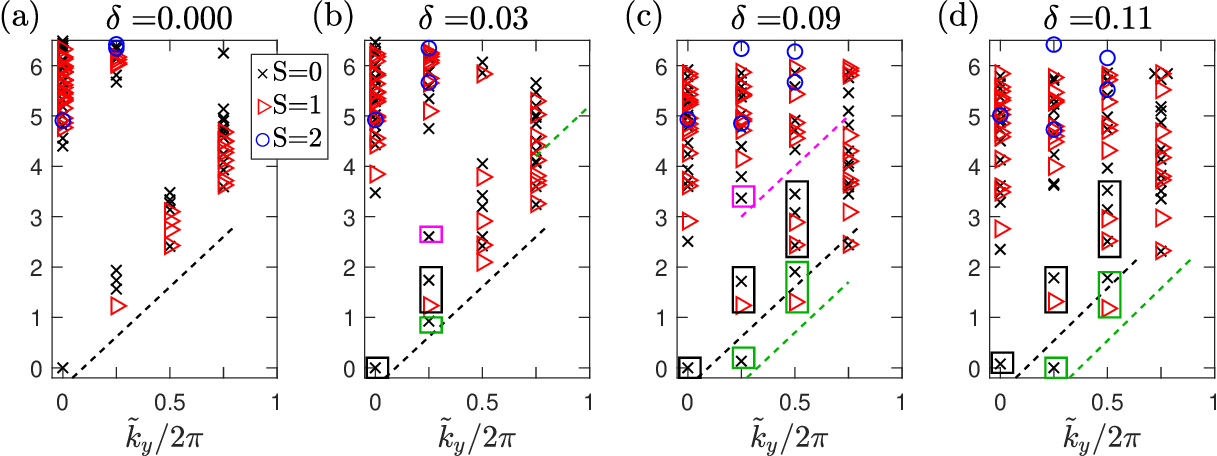}
\caption{{\bf Entanglement spectrum of doped superconducting states at $U=4$ on a cylinder of width $L_y=8$.} 
The green and pink dashed lines indicate the subleading branches of the Chern insulator (corresponding to $Q=\pm2$ branches in Fig.~\ref{fig:compare_CI_CSL}(c)) that evolve towards lower and higher energies, respectively.
Data are shown for $(D,\chi_{\mathrm{iPESS}})=(8,60)$.}
\label{fig:SC_ES_U4}
\end{localgraphicspath}
\end{figure}

{\bf \emph{Common features of the entanglement spectrum.}}~
Previous studies on finite cylinders have demonstrated that doping on both sides of the Mott transition yields superconductivity with the same global topological spin response~\cite{chen2025topological,kuhlenkamp2025robust}. However, whether the superconducting states emerging from the Chern insulator and chiral spin liquid parent phases are identical beyond such a global property remains uninvestigated. To shed light on this question, we examine the entanglement spectrum of the doped superconducting states across the Mott transition.

The ES for the doped Chern insulator is presented in Fig.~\ref{fig:SC_ES_U4} for $U=4$ on a cylinder of width $L_y=8$. Due to the two-fold Brillouin zone folding induced by pairing order, the subleading $Q=\pm2$ branches associated with charge fluctuations in Fig.~\ref{fig:compare_CI_CSL}(c) appear at $\tilde{k}_y/2\pi=1/4$  in Fig.~\ref{fig:SC_ES_U4}(a). This momentum shift can be understood as arising from adding or removing two particles (or holes) at the single-particle mode (edge mode) closest to zero energy, with both spin components simultaneously occupied or emptied (see SM for details).
With increasing $\delta$ and in the presence of pairing order that breaks charge $U(1)$ symmetry to $Z_2$, the two degenerate subleading branches split and evolve toward lower and higher energies, respectively, indicating a pronounced particle-hole asymmetry. The downward-moving branch reaches energies comparable to the leading branch at $\tilde{k}_y=0$ for $\delta\sim0.1$, exhibiting the same level counting as in the Chern insulating phase [Fig.~\ref{fig:SC_ES_U4} (b)-(d)]. 

As already discussed, similar low-energy branches also emerge on the CSL side [Fig.~\ref{fig:SC_ES}], indicating additional low-lying entanglement spectrum branches common to the doped superconducting states. We note that the corresponding branch appears at $\tilde{k}_y/2\pi=3/4$ on the same $L_y=8$ cylinder for $U>U_{c_1}$, which cannot be accounted for within the above non-interacting picture in terms of filling the single-particle edge mode. 
The different momenta of the low-lying entanglement-spectrum branches may reflect distinct nonuniversal excitation or edge-spectrum structures inherited from the Chern insulator and CSL parent states.
The microscopic origin of this momentum shift remains to be understood.
Overall, these results show that the entanglement spectrum retains the chiral dispersion and level counting of the parent topological phases while developing additional low-energy branches upon doping.

{\bf \emph{Conclusion.}}~
In summary, by advancing a simplex tensor network approach, we provide thermodynamic-limit evidence for chiral superconductivity emerging from doping chiral topological phases in a repulsive fermionic system, as demonstrated in the triangular Hofstadter-Hubbard model by a finite complex pairing order parameter in the infinite-size limit.
Starting from robust chiral parent phases that include both the Chern insulator and the chiral spin liquid, hole doping induces a uniform chiral superconducting ground state over a broad interaction-doping regime. 
Beyond establishing superconducting long-range order, we further analyze the structure of the superconducting state. We uncover a distinct pocket with reversed phase winding near the Mott criticality, revealing subtle competition within the chiral superconducting state. In addition, we find that the entanglement spectrum retains the chiral dispersion and level counting of the parent topological phases while developing additional low-energy branches upon doping. 
These results establish that chiral superconductivity emerges robustly from doped chiral topological phases in the thermodynamic limit.

The present work establishes thermodynamic-limit chiral superconductivity in the triangular Hofstadter-Hubbard model as a concrete microscopic realization with explicit magnetic flux and robust chiral topological parent states. An important open direction is to understand how the phase diagram evolves in other lattice geometries and in systems with different types of chiral topological order.

The infinite tensor network framework developed here provides a natural route to address this question in other settings. Promising directions include: (i) extending the Hofstadter-Hubbard model to different lattice geometries such as square or kagome lattices, as well as to $\rm{SU}(N)$ generalizations with multiple fermionic flavors~\cite{zhang2026phases}; (ii) investigating systems with spontaneously chiral topological order, such as the triangular-lattice Hubbard model~\cite{szasz2020chiral} or kagome $t_1-t_2-t_3$ Hubbard models~\cite{gong2014emergent}; and (iii) exploring doped fractional Chern insulators, where recent advances in fermionic tensor network constructions make such studies increasingly feasible~\cite{chenhao2025simulating}.
A unifying goal of these directions is to establish, at the thermodynamic limit, robust evidence for superconductivity emerging from doping chiral topological phases of fermions with repulsive interactions.

{\bf \emph{Acknowledgment}}.~Sen Niu is thankful for insightful discussion with Didier Poilblanc, Leon Balents, Rong-Yang Sun, Jheng-Wei Li, Wen-Yuan Liu, Rui-Zhen Huang, Juraj Hasik and Lin Zhang. The research  was supported by the U.S. Department of
Energy, Office of Basic Energy Sciences under Grant No. DE-FG02-06ER46305 (SN, DNS) for numerical study of strongly correlated topological phases. 
This work is also supported by the US National Science Foundation (NSF) Grant No. PHY-2216774 (YP). 
The numerical simulation is partially done  through the support of the  NSF instrument grant  DMR-2406524, 
and TensorKit.jl~\cite{devos2025tensorkit} and YASTN~\cite{rams2025yastn} packages are used for automatic differentiation with $\rm{SU}(2)$ and $Z_2$ symmetric tensors, respectively.

\clearpage
\appendix

\onecolumngrid

\begin{center}
{\Large Supplemental Materials}\\[0.5em]
\end{center}

\vspace{1em}

\tableofcontents

\noindent
This Supplemental Material provides technical details, additional numerical evidence, and algorithmic benchmarks that support and complement the results presented in the main text. The material is organized as follows. Section~I introduces the fermionic iPESS formalism and numerical procedures underlying our simulations. Section~II presents additional numerical data that corroborate the physical conclusions. Section~III benchmarks different tensor-network ans\"atze and optimization strategies, clarifying the advantages of the variational fermionic iPESS approach adopted in this work.

\section{Methods and technical details}
This section summarizes the fermionic iPESS framework and the numerical techniques employed throughout this work. We describe the tensor structure, fermionic sign conventions, entanglement-spectrum calculations, symmetry implementation, and gauge choices that underpin all results in the main text.

\subsection{Fermionic tensors for wavefunction and Hamiltonian}

We introduce the fermionic iPESS ans\"atz and its tensor structure, emphasizing fermion-parity conservation and the exact treatment of fermionic statistics without introducing nonlocal Jordan--Wigner strings.

The fermionic iPESS can be viewed as a generalization of the fermionic iPEPS to the triangular lattice. Its memory cost is lower than that of iPEPS, since the elementary $\hat{S}$ and $\hat{T}$ tensors each carry only three virtual indices. Below we briefly review the fermionic tensor formalism~\cite{barthel2009contraction} and derive the corresponding swap-gate structure relevant for fermionic iPESS.

In our fermionic iPESS ans\"atz, the wave-function tensors take the forms
$\hat{T}=T_{LUM}|LU\rangle\langle M|$ and
$\hat{S}=S_{MDR}^{p}|Mp\rangle\langle DR|$,
where $L,D,R,U,M$ and $p$ denote virtual and physical fermionic degrees of freedom, respectively. Throughout this section, these labels also encode the corresponding fermion parity quantum numbers of the corresponding legs/indices. Fermionic exchange obeys
$|pq\rangle =(-1)^{pq}|qp\rangle$ and
$|p\rangle \langle q| =(-1)^{pq}\langle q| |p\rangle$,
where bras and kets label outgoing and incoming tensor indices, respectively.
The Hermitian conjugation satisfies
$(|pq\rangle)^{\dagger}=(|p\rangle|q\rangle)^{\dagger}=\langle q|\langle p|=\langle pq|$.

A key feature of fermionic tensors is the conservation of $Z_2$ fermion parity, which imposes the constraints
\begin{align}
\mod(L+U-M,2)&=0, \notag\\
\mod(M+D+R-p,2)&=0,
\end{align}
for nonvanishing elements of $\hat{T}$ and $\hat{S}$, respectively.
When $\mathrm{SU}(2)$ symmetry is imposed, fermion parity follows directly from the spin representation: an even (odd) number of fermions corresponds to an integer (half-integer) total spin.

Fermionic hopping operators $c_{1\sigma}^{\dagger}c_{2\sigma}$ are decomposed into $Z_2$-parity-conserving tensors,
\begin{equation}
\includegraphics[scale = 0.6]{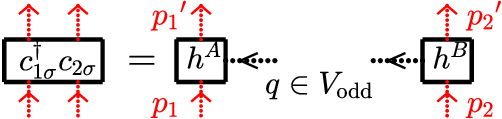}, \label{eq:ipepstensor}
\end{equation}
where the auxiliary virtual index $q$ is fixed to be fermion-parity odd, reflecting the change of fermion number by one. 
Therefore, the parity-conserving $\hat{h}^A=h^{A}_{p_1',p_1,q}|p_1'\rangle\langle p_1|\langle q|$ and $\hat{h}^B=h^{B}_{q,p_2',p_2}|qp_2'\rangle\langle p_2|$ operators commute with local iPESS $\hat{S}, \hat{T}$ tensors, and only local swap gates need to be accounted for during contractions. Long-range Jordan-Wigner strings are therefore avoided. 

Expectation values of observables are evaluated by applying local operators to the original $\hat{S}$ and $\hat{T}$ tensors, yielding modified tensors $\hat{S}'$ and $\hat{T}'$. For single-site parity conserving operators such as occupation number or double occupancy, the operator has the form $\hat{O}=O_{p',p}|p'\rangle \langle p|$,  the action $\hat{O}_r$ on position $r$ yields the following tensor diagram:
\begin{equation}
\includegraphics[scale = 0.3]{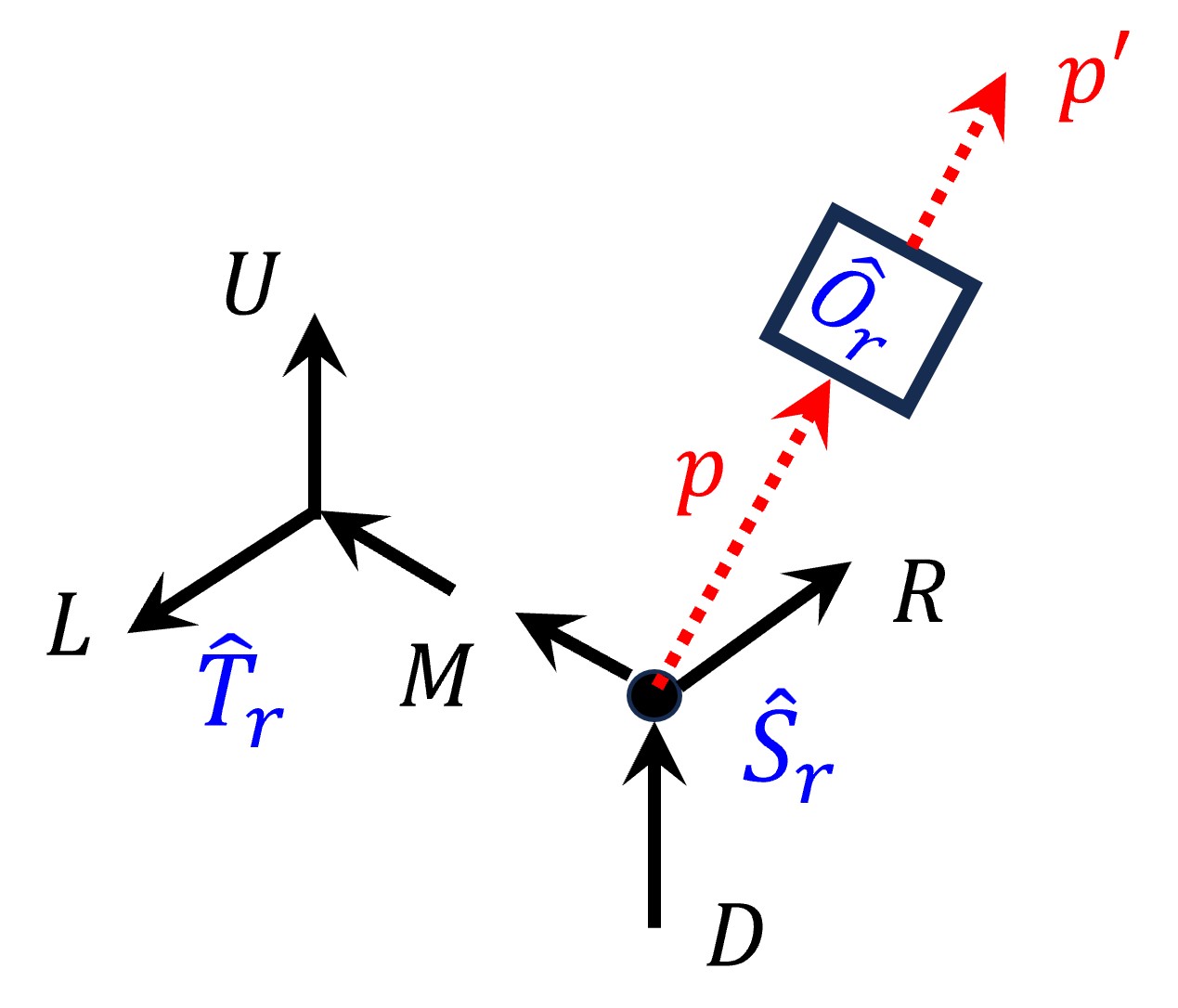} \label{eq:ipepstensor}
\end{equation}
The corresponding modified tensors on site $r$ can be written as
\begin{align}
\hat{T}_r & \rightarrow \hat{T}_r, \notag\\
\hat{S}_r & \rightarrow \hat{S}_r^{'}=\hat{O}_r\hat{S}_r  =O_{p_r',p_r}S_{MDR}^{p_r}|M p_r'\rangle\langle DR|.
\end{align}

For electron hopping operators, extra fermionic signs emerge. Let's consider the hopping along x-axis, where $\hat{h}^A$ acts on site $r$ and $\hat{h}^B$ acts on site $r+e_1$. The tensor diagram for this operation can be expressed as
\begin{equation}
\includegraphics[scale = 0.4]{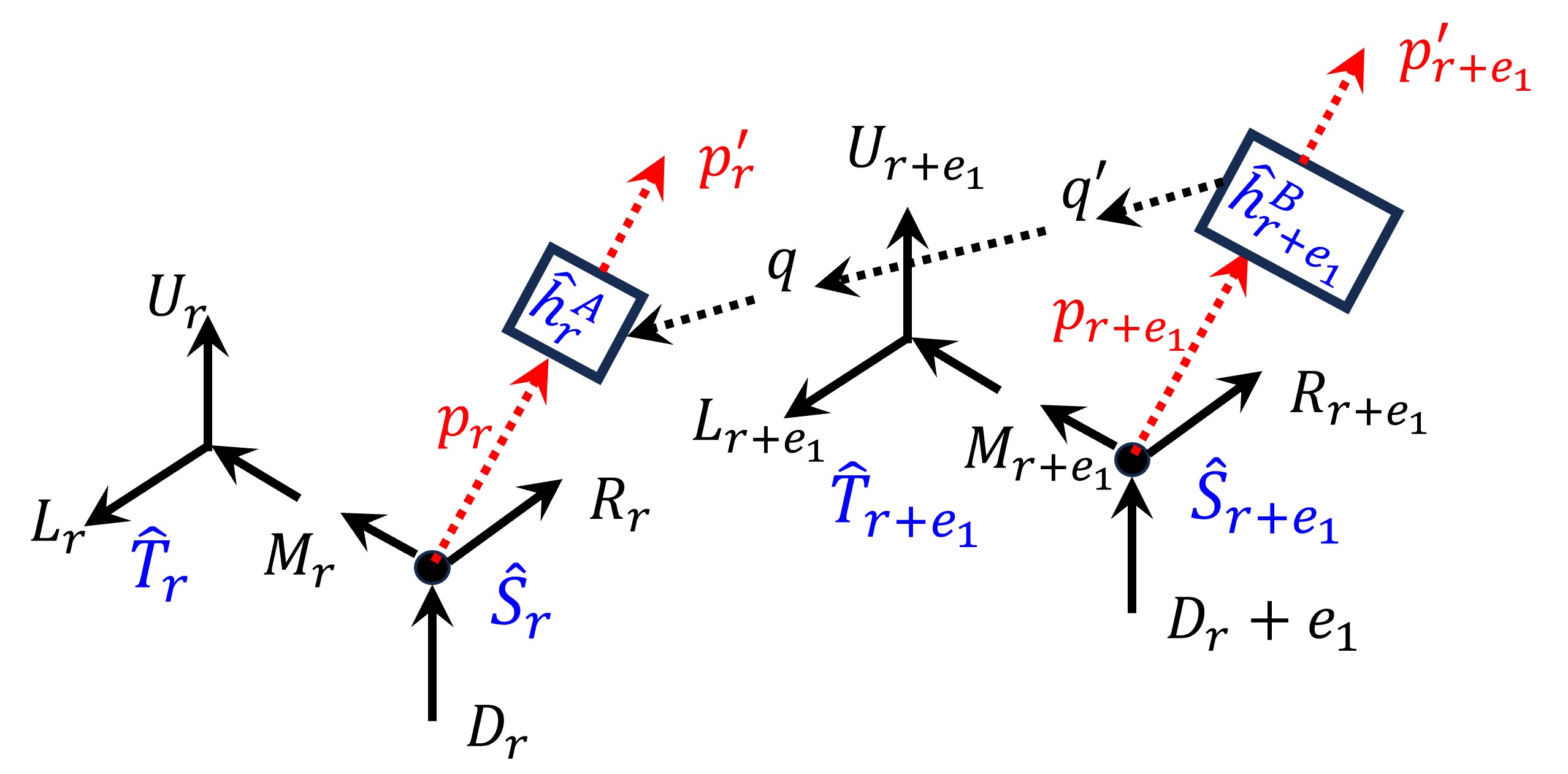} \label{eq:ipepstensor}
\end{equation}
For site $r$, the corresponding modified tensors take the form
\begin{align}
\hat{T}_r & \rightarrow \hat{T}_r, \notag\\
\hat{S}_r & \rightarrow \hat{S}_r^{'}=\hat{h}^A_r\hat{S}_r  = (h^{A}_{p_r',p_r,q}|p_r'\rangle\langle p_r|\langle q|)( S_{MDR}^{p_r}|M p_r\rangle\langle DR|)\notag\\
&=(-1)^{p_r(q+M)}h^{A}_{p_r',p_r,q} S_{MDR}^{p_r}|p_r'\rangle\langle q| |M\rangle\langle DR|\notag\\
&=(-1)^{p_r(q+M)}(-1)^{q(M+R)}(-1)^{p_r'M}h^{A}_{p_r',p_r,q} S_{MDR}^{p_r} |Mp_r'\rangle\langle qR|\langle D|,
\end{align}
where $\langle R|\langle q|=\langle qR|$ can be viewed as the extended right leg by absorbing the virtual $q$ leg from the operator.
For site $r+e_1$, the corresponding modified tensors take the form
\begin{align}
\hat{T}_{r+e_1} & \rightarrow \hat{T}_{r+e_1}'= \hat{I}\hat{T}_{r+e_1}=(\delta_{q,q'}|q\rangle \langle q'|)(T_{LUM}|LU\rangle\langle M|) \notag\\
&=\delta_{q,q'}T_{LUM}|qLU\rangle\langle q'M| \notag\\ 
\hat{S}_{r+e_1} & \rightarrow \hat{S}_{r+e_1}^{'}=\hat{h}^B_{r+e_1}\hat{S}_{r+e_1}  = (h^{B}_{q'p_{r+e_1}',p_{r+e_1},q}|q'p_{r+e_1}'\rangle\langle p_{r+e_1}|)( S_{MDR}^{p_{r+e_1}}|M p_{r+e_1}\rangle\langle DR|)\notag\\
&=(-1)^{(p_{r+e_1}'+p_{r+e_1})M}h^{B}_{q'p_{r+e_1}',p_{r+e_1},q}S_{MDR}^{p_{r+e_1}}|q'Mp_{r+e_1}' \rangle\langle DR|),
\end{align}
where $\hat{I}$ is the identity that is absorbed by $\hat{T}_{r+e_1}$, $|qL\rangle$ can be viewed as the extended left leg, and $|q'M\rangle$ can be viewed as the extended middle leg. The $\langle qR|$ leg on site $r$ will be contracted with the $|qL\rangle$ leg on site $r+e_1$.
The hopping along the other two directions (lattice vectors) $e_2$, $e_1+e_2$ can be derived in similar ways.

To evaluate wavefunction norms and observables, we employ the CTMRG method to contract the state on an infinite lattice.  
While one may directly combine $\hat{S}$ and $\hat{T}$ tensors into iPEPS tensors via $\hat{A}_r=\hat{T}_r\hat{S}_r=T_{LUM}S_{MDR}^{p_r}|LUp_r\rangle\langle DR|$, retaining the iPESS structure significantly reduces memory requirements.
To calculate wavefunction overlap, we first contract physical indices by constructing fermionic double-layer tensors and permute the virtual indices in the following order:
\begin{align}
\hat{T}^{\dagger}\hat{T} & =T_{L'U'M'}^{*}T_{LUM}|M'\rangle\langle U'|\langle L'||L\rangle|U\rangle\langle M| \notag\\
 & =T_{L'U'M'}^{*}T_{LUM}(-1)^{U'(L'+L)}(-1)^{M'(L+L'+U+U'+M)}\langle L'||L\rangle\langle U'||U\rangle\langle M||M'\rangle \notag\\
 & =T_{L'U'M'}^{*}T_{LUM}(-1)^{U'(L'+L)+M'(L'+U')}\langle L'||L\rangle\langle U'||U\rangle\langle M||M'\rangle \notag\\
 \hat{S}^{\dagger}\hat{S} & =(S_{M'D'R'}^{p})^{*}S_{MDR}^{p}|D'\rangle|R'\rangle\langle p|\langle M'||M\rangle|p\rangle\langle R|\langle D| \notag\\
 & =(S_{M'D'R'}^{p})^{*}S_{MDR}^{p}(-1)^{pM+pM'}(-1)^{R'(M+M'+R)}(-1)^{D'(M+M'+R+R'+D)}\langle M'||M\rangle\langle R||R'\rangle\langle D||D'\rangle
\end{align}
These $\langle L'||L\rangle, \langle D||D'\rangle, \langle R||R'\rangle,\langle U'||U\rangle, \langle M||M'\rangle$ legs correspond to the fused legs of double layer tensors. To trace the intermediate $M,M'$ indices connecting $\hat{S}, \hat{T}$ tensors, we have used
\begin{align}
\text{Tr}_{M'}[|M'\rangle\langle M'|]=\text{Tr}_{M'}[(-1)^{M'}\langle M'|M'\rangle]=(-1)^{M'}.
\end{align}
The fermionic signs discussed above correspond to local swap gates that must be included in fermionic iPESS, in contrast to the bosonic case. The additional computational cost is, however, minor compared to bosonic iPESS. Once the physical indices are contracted into the double-layer tensors, the virtual indices are approximately contracted using the standard CTMRG method. The accuracy of CTMRG is controlled by the environment dimension $\chi_{\rm iPESS}$, and typically $\chi_{\rm iPESS} \approx D^2$ is sufficient to accurately evaluate the energy of a state with bond dimension $D$.

A major bottleneck in large-$D$ simulations is memory usage. For automatic differentiation of an $m \times n$ cell (fermionic) iPEPS, one typically needs to store $m \times n$ rank-4 double-layer iPEPS tensors. These tensors store not only the variational parameters but also the intermediates needed to evaluate observables, leading to a leading memory cost of $O(m n D^8)$. In contrast, for (fermionic) iPESS, only $m \times n$ rank-3 double-layer tensors $\hat{T}^\dagger \hat{T}$ and $\hat{S}^\dagger \hat{S}$ are needed to store the variational parameters, which reduces the memory cost to $O(m n D^6)$.
During CTMRG iterations or when evaluating Hamiltonian terms, it is necessary to combine iPESS tensors into rank-4 double-layer iPEPS tensors $\hat{A}^\dagger \hat{A} = \hat{T}^\dagger \hat{T}  \hat{S}^\dagger \hat{S}$. The memory cost of this step can be reduced from $O(m n D^8)$ to $O(D^8)$ using the checkpointing technique~\cite{liao2019differentiable}. Consequently, the total memory cost for iPESS is $O(m n D^6) + O(D^8)$, allowing gradient-based optimization at larger bond dimensions compared to iPEPS, especially for sufficiently large unit cells (e.g., $6 \times 6$ for N\'eel order and $4\times 4$ for SC order).

\subsection{Calculation of entanglement spectrum }
The entanglement spectrum (ES) of $1\times1$ unit-cell iPEPS/iPESS on finite-width cylinders can be computed exactly using exact contraction methods~\cite{cirac2011entanglement,poilblanc2012topological}. By employing CTMRG environments as approximate fixed points of the transfer matrix~\cite{poilblanc20162}, one can effectively access cylinders with twice the original width. Here, we generalize this CTMRG-based approach to fermionic systems with generic $m\times n$ unit cells.

\begin{figure}[h]
\begin{localgraphicspath}{{figs/draft_fig/}}
\includegraphics[width=0.8\columnwidth]{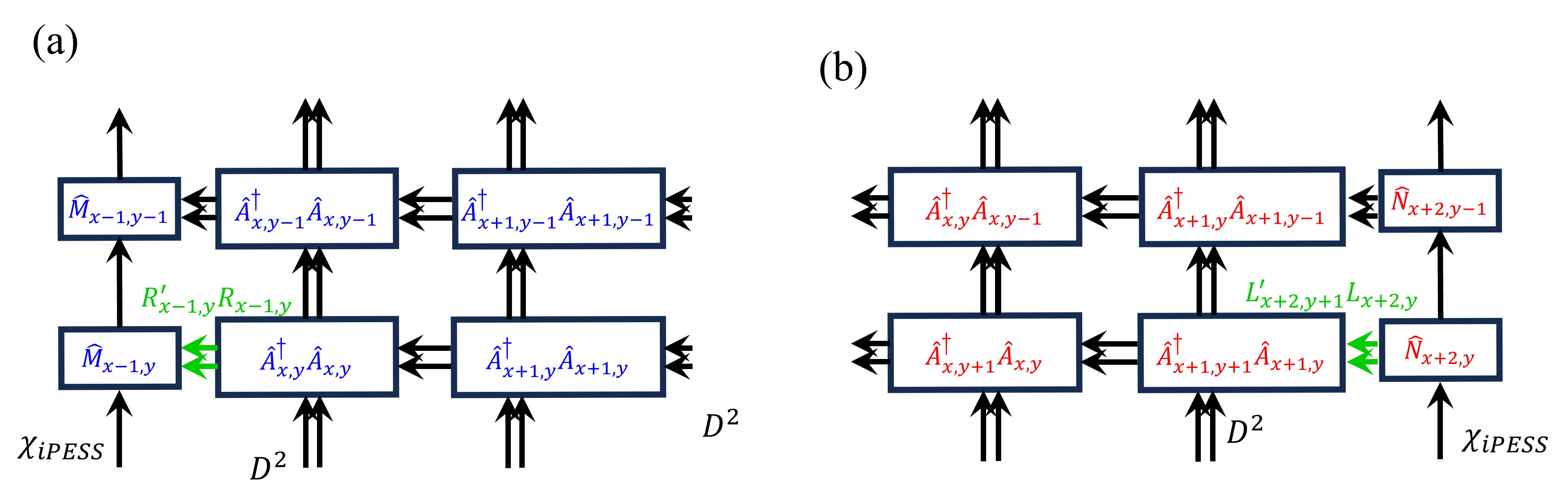}
\caption{{\bf Double layer tensors and boundary CTMRG environments for calculating entanglement spectrum.} A generic unit-cell is assumed. (a) Standard double layer tensors for part of a unit-cell and their left CTMRG environments. (b) Twisted double layer tensors along $y$ direction by one site and their right CTMRG environments.}
\label{fig:CMT_envs}
\end{localgraphicspath}
\end{figure}

We map triangular lattice sites $(x\mathbf{e}_1,y\mathbf{e}_2)$ onto square-lattice coordinates $(x,y)$ and consider an infinitely long cylinder along the $x$ direction with finite circumference $L_y$ along $y$. Since the Hamiltonian (in the original gauge) is translationally invariant along $y$, the variational state is expected to be an approximate eigenstate of the translation operator $T_y$.
A key limitation of earlier approaches~\cite{cirac2011entanglement,poilblanc2012topological,poilblanc20162} is that the translation operator $T_y$ acts on virtual legs, which requires all virtual $\mathrm{SU}(2)$ bonds to be identical. This condition is generally violated for states obtained from imaginary-time evolution (either full or simple update), especially for larger unit cells.
Inspired by the DMRG-based ES algorithm~\cite{cincio2013characterizing}, we instead compute momentum-resolved ES using standard and twisted transfer matrices on the cylinder, where the translation operator $T_y$ acts on the physical legs rather than the virtual ones. The transfer matrices are constructed from double-layer tensors~\cite{cirac2011entanglement}. Specifically, standard double-layer tensors form the two-dimensional tensor network obtained by contracting physical legs between $\langle\psi|$ and $|\psi\rangle$ [Fig.~\ref{fig:CMT_envs}(a)], while twisted double-layer tensors are formed by contracting physical legs between $\langle\psi|$ and $T_y|\psi\rangle$ [Fig.~\ref{fig:CMT_envs}(b)]. We perform CTMRG separately for these two networks to obtain the corresponding coarse-grained boundary environments.

\begin{figure}[h]
\begin{localgraphicspath}{{figs/draft_fig/}}
\includegraphics[width=0.4\columnwidth]{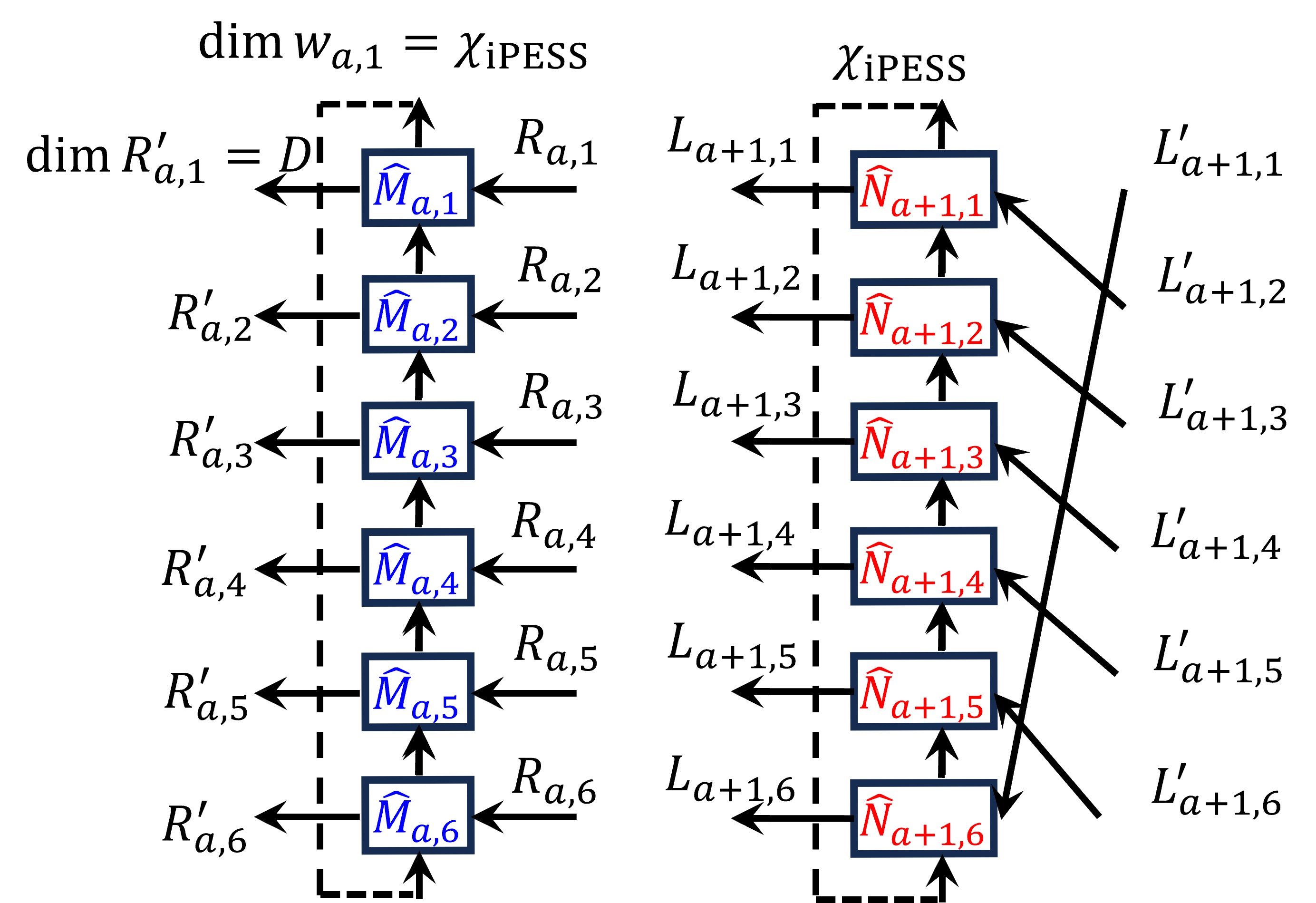}
\caption{{\bf CTMRG environments as transfer matrix fixed points on the width-$6$ cylinder.}  Left column: Environment tensors (boundary matrix product state) from standard double layer tensors in Fig.~\ref{fig:CMT_envs} (a). Right column: Environment tensors from twisted double layer tensors in Fig.~\ref{fig:CMT_envs} (b).}
\label{fig:ES_diagram}
\end{localgraphicspath}
\end{figure}

The transfer-matrix fixed point on a cylinder can be approximated by a periodic matrix product state (MPS) obtained by wrapping CTMRG boundary environments around the circumference~\cite{poilblanc20162}. We illustrate the procedure for a width-$6$ cylinder in Fig.~\ref{fig:ES_diagram}. Let the left boundary environments of the standard (untwisted) transfer matrix be denoted by
$\hat{M}^{a,b}=M_{w_{a,b},R_{a,b},R_{a,b}',w_{a,b+1}}^{a,b} |w_{a,b} \rangle\langle R_{a,b}||R_{a,b}'\rangle \langle w_{a,b+1}|$ 
at position $(a,b)$. Due to the $m\times n$ periodicity of the iPESS ans\"atz, we have
$M^{a,b}=M^{\mathrm{mod}(a,m),\mathrm{mod}(b,n)}$.
The vertical index $w$ has dimension $\chi_{\rm iPESS}$, while the horizontal indices $R,R'$ have dimension $D$, the iPESS bond dimension.
The left fixed point $\hat{\sigma}_L$ is obtained by tracing over the vertical indices around the cylinder,
\begin{align}
\hat{\sigma}_L &=\text{Tr}_{w_{a,1},w_{a,2},w_{a,3},w_{a,4},w_{a,5},w_{a,6}}[\hat{M}^{a,1}\hat{M}^{a,2}\hat{M}^{a,3}\hat{M}^{a,4}\hat{M}^{a,5}\hat{M}^{a,6}] \notag\\
&= \text{Tr}_{w_{a,1}} [\delta_{w_{a,1},w_{a,7}}M_{w_{a,1},R_{a,1},R_{a,1}',w_{a,2}}^{a,1} M_{w_{a,2},R_{a,2},R_{a,2}',w_{a,3}}^{a,2} M_{w_{a,3},R_{a,3},R_{a,3}',w_{a,4}}^{a,3} \notag \\
&\times M_{w_{a,4},R_{a,4},R_{a,4}',w_{a,5}}^{a,4} M_{w_{a,5},R_{a,5},R_{a,5}',w_{a,6}}^{a,5} M_{w_{a,6},R_{a,6},R_{a,6}',w_{a,7}}^{a,6} \notag\\
&\times|w_{a,1} \rangle\langle R_{a,1}||R_{a,1}'\rangle  \langle R_{a,2}||R_{a,2}'\rangle \langle R_{a,3}||R_{a,3}'\rangle  \langle R_{a,4}||R_{a,4}'\rangle \langle R_{a,5}||R_{a,5}'\rangle \langle R_{a,6}||R_{a,6}'\rangle \langle w_{a,7}|] \notag\\
&=(-1)^{w_{a,1}}M_{w_{a,1},R_{a,1},R_{a,1}',w_{a,2}}^{a,1} M_{w_{a,2},R_{a,2},R_{a,2}',w_{a,3}}^{a,2} M_{w_{a,3},R_{a,3},R_{a,3}',w_{a,4}}^{a,3} \notag \\
&\times M_{w_{a,4},R_{a,4},R_{a,4}',w_{a,5}}^{a,4} M_{w_{a,5},R_{a,5},R_{a,5}',w_{a,6}}^{a,5} M_{w_{a,6},R_{a,6},R_{a,6}',w_{a,1}}^{a,6} \notag\\
&\times\langle R_{a,1}||R_{a,1}'\rangle  \langle R_{a,2}||R_{a,2}'\rangle \langle R_{a,3}||R_{a,3}'\rangle \langle R_{a,4}||R_{a,4}'\rangle \langle R_{a,5}||R_{a,5}'\rangle \langle R_{a,6}||R_{a,6}'\rangle,  
\label{eq:left_sigma1}
\end{align}
which should be rearranged into a matrix acting on the virtual bond space. The fermionic sign structure arising from index permutations is handled explicitly, leading to the matrix form
\begin{align}
& \langle R_{a,1}||R_{a,1}'\rangle  \langle R_{a,2}||R_{a,2}'\rangle \langle R_{a,3}||R_{a,3}'\rangle \langle R_{a,4}||R_{a,4}'\rangle \langle R_{a,5}||R_{a,5}'\rangle \langle R_{a,6}||R_{a,6}'\rangle \notag\\
=& (-1)^{R_{a,6}'(R_{a,6}+\sum_{b=1}^{5}(R_{a,b}+R_{a,b}'))}|R_{a,6}'\rangle \langle R_{a,1}||R_{a,1}'\rangle  \langle R_{a,2}||R_{a,2}'\rangle \langle R_{a,3}||R_{a,3}'\rangle \langle R_{a,4}||R_{a,4}'\rangle \langle R_{a,5}||R_{a,5}'\rangle \langle R_{a,6}| \notag\\
=& (-1)^{R_{a,6}'(R_{a,6}+w_{a,1}+w_{a,6})}|R_{a,6}'\rangle \langle R_{a,1}||R_{a,1}'\rangle  \langle R_{a,2}||R_{a,2}'\rangle \langle R_{a,3}||R_{a,3}'\rangle \langle R_{a,4}||R_{a,4}'\rangle \langle R_{a,5}||R_{a,5}'\rangle \langle R_{a,6}| \notag\\
=&\cdot\cdot\cdot\cdot \notag\\
=& (-1)^{\sum_{b=1}^{6}R_{a,b}'R_{a,b}+\sum_{b=1}^{6}R_{a,b}'w_{a,1}+\sum_{b=1}^{5}R_{a,b}'w_{a,b}}|R_{a,6}'\rangle |R_{a,5}'\rangle |R_{a,4}'\rangle |R_{a,3}'\rangle |R_{a,2}'\rangle |R_{a,1}'\rangle \langle R_{a,1}|  \langle R_{a,2}|\langle R_{a,3}| \langle R_{a,4}| \langle R_{a,5}| \langle R_{a,6}|\notag\\
=& (-1)^{\sum_{b=1}^{6}R_{a,b}'R_{a,b}+\sum_{b=1}^{6}R_{a,b}'w_{a,1}+\sum_{b=1}^{5}R_{a,b}'w_{a,b}}|R_{a,1}' R_{a,2}' R_{a,3}' R_{a,4}' R_{a,5}' R_{a,6}'\rangle \langle R_{a,1} R_{a,2} R_{a,3} R_{a,4} R_{a,5} R_{a,6}|.
\label{eq:left_sigma2}
\end{align}

Similarly, the right boundary environments take the form $\hat{N}^{a+1,b}=N_{w_{a+1,b},L_{a+1,b}',L_{a+1,b},w_{a+1,b+1}}^{a+1,b} |w_{a+q,b} \rangle$ $\langle L_{a+1,b+1}'|| L_{a+1,b}\rangle \langle w_{a+1,b+1}|$ at position $a+1,b$, where the distinction between coordinates of $L_{a+1,b+1}'$ and $L_{a+1,b}$ originates from the $T_y$ shift of the $A^{\dagger}$ layer tensors [Fig.~\ref{fig:CMT_envs} (b)].  We also transform the right fixed point of $\hat{\sigma}_R$ to the matrix form $|L_{a+1,1} L_{a+1,2} L_{a+1,3} L_{a+1,4} L_{a+1,5} L_{a+1,6}\rangle \langle L_{a+1,1}' L_{a+1,2}' L_{a+1,3}' L_{a+1,4}' L_{a+1,5}' L_{a+1,6}'|$  following the same procedure.
The entanglement spectrum is then obtained from the eigenvalues of the effective reduced density matrix defined on the virtual bonds~\cite{cirac2011entanglement},
\begin{align}
\rho=\hat{\sigma}_L\hat{\sigma}_R,
\end{align}
where $\hat{\sigma}_L$ and $\hat{\sigma}_R$ live on nearest neighbour columns and their contiguous legs satisfy the orthonormal condition
\begin{align}
\langle  R_{a,1} R_{a,2} R_{a,3} R_{a,4} R_{a,5} R_{a,6}|L_{a+1,1} L_{a+1,2} L_{a+1,3} L_{a+1,4} L_{a+1,5} L_{a+1,6}\rangle=\prod_{b=1}^{6}\delta_{R_{a,b},L_{a+1,b}}.
\end{align}
The phases of the eigenvalues yield the momentum quantum numbers via $e^{i k_y}$. In the CI and CSL phases, despite using a $2\times2$ unit cell for iPESS, we find that $k_y$ remains nearly perfectly quantized, indicating that physical-site translation symmetry $T_y$ is well preserved.

For doped superconducting (SC) states, the pairing field is staggered along the $y$ direction, and the optimized state is only an eigenstate of the two-site translation operator $T_y^2$. In this case, the entanglement spectrum can be computed by a straightforward modification of the above procedure.  

\subsection{$\rm{SU}(2)$ symmetric virtual space}
The structure of the $\mathrm{SU}(2)$ virtual bond is not uniquely fixed for a given bond dimension $D$. When the full $\mathrm{U}(1)\times\mathrm{SU}(2)$ symmetry is imposed, the physical Hilbert space can be written as the direct sum
$\mathcal{V}_p=(0,0)\oplus(1,1/2)\oplus(2,0)$, 
where the first quantum number denotes the $\mathrm{U}(1)$ particle number and the second labels the $\mathrm{SU}(2)$ spin irreducible representation (IRREP). In our simulations, however, we work in the grand-canonical ensemble and therefore enforce only $\mathrm{SU}(2)$ symmetry. The physical space is then expressed as
$\mathcal{V}_p=2*(0)\oplus 1*(1/2)$, where the prefactors indicate the degeneracies of the corresponding IRREPs. 
A natural choice for the virtual bond space is $\mathcal{V}_v=\mathcal{V}_p$, which corresponds to bond dimension $D=4$. In practice, however, we find that $D=4$ is insufficient to accurately describe the Hofstadter-Hubbard model, and larger bond dimensions are required to achieve lower variational energies. Among the possible choices of virtual spaces at moderate bond dimensions, we find that
$\mathcal{V}_v=2*(0)\oplus 2*(1/2)$ for $D=6$ and $\mathcal{V}_v=4*(0)\oplus 2*(1/2)$ for $D=8$ 
yield the lowest energies within our variational optimization method.

As the number of admissible virtual-bond choices grows rapidly with increasing $D$, we further employ a full-update method, which performs imaginary-time evolution starting from the variationally optimized states to effectively enlarge the virtual space and further reduce the energy. In contrast to purely variational optimization, the energy in the full-update procedure is generally not a monotonic function of the total imaginary time $\tau$: it typically decreases for $0<\tau<\tau_c$ and increases for $\tau>\tau_c$. We therefore monitor the energy at each imaginary-time step $d\tau$ and terminate the evolution once the energy ceases to decrease. The combined CTMRG-based variational optimization and full-update approach enables accurate simulations, as evidenced by the energy benchmarks shown in Fig.~2(a)-(b) of the main text.

\subsection{Gauges and gauge transformation of the $\pi/2$-flux Hofstadter Hamiltonian}
In most of our numerical simulations, we adopt the original gauge $t_{\boldsymbol{r},\boldsymbol{r}+\boldsymbol{e}_1}=it$, $t_{\boldsymbol{r},\boldsymbol{r}+\boldsymbol{e}_2}=(-1)^{x}t$, $t_{\boldsymbol{r},\boldsymbol{r}+\boldsymbol{e}_3}=(-1)^{x-1}t$ following Ref.~\cite{kuhlenkamp2024chiral}. 
In this gauge, the Hamiltonian has a $2\times1$ unit cell and preserves single-site translation symmetry along the $y$ direction ($T_y$). As a result, the ES of symmetry-unbroken CI and CSL phases are naturally labeled by the momentum $k_y$.

For direct comparison with the pairing symmetries reported in previous DMRG studies~\cite{divic2025anyon,chen2025topological,kuhlenkamp2025robust}, we further consider the following gauge transformation:
\begin{align}
c_{2x-1,2y-1}^{\dagger}&	\rightarrow(-1)^{x}c_{2x-1,2y-1}^{\dagger},\notag \\
c_{2x,2y-1}^{\dagger}&	\rightarrow(-1)^{x}c_{2x,2y-1}^{\dagger},\notag \\
c_{2x-1,2y}^{\dagger}&	\rightarrow i(-1)^{x}c_{2x-1,2y}^{\dagger},\notag \\
c_{2x,2y}^{\dagger}&	\rightarrow i(-1)^{x}c_{2x,2y}^{\dagger},
\end{align}
which maps the Hamiltonian to a $C_6$-symmetric gauge with a $2\times2$ unit cell. All SC pairing patterns shown in this work are presented in this $C_6$-symmetric gauge.

\section{Additional numerical data}

\begin{figure}[h]
\begin{localgraphicspath}{{figs/draft_fig/}}
\includegraphics[width=0.4\columnwidth]{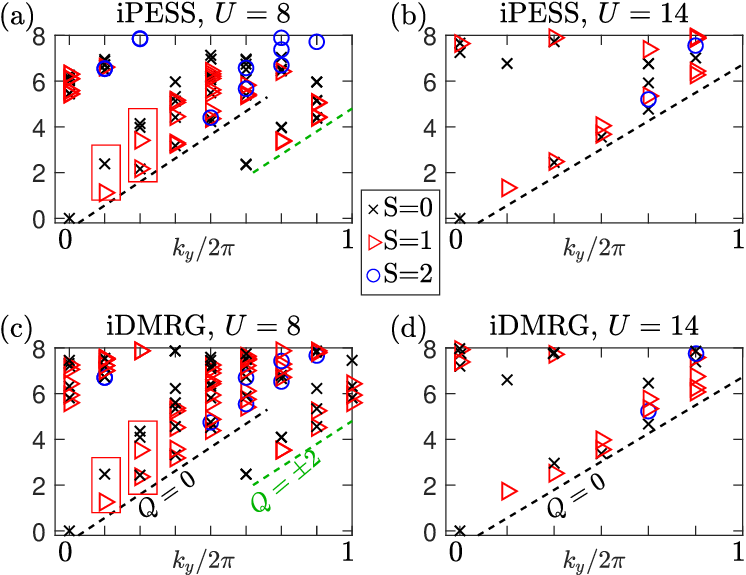}
\caption{{\bf Comparison between variational iPESS and iDMRG entanglement spectrum at half filling.} (a)-(b) iPESS with $(D,\chi_{\rm{iPESS}})=(12,60)$. (c)-(d) iDMRG with $D_{\rm{iDMRG}}=16000$. Antiperiodic boundary condition is used for $L_y=8$, and periodic boundary condition is used for $L_y=6$.}
\label{fig:SM_ES_DMRG}
\end{localgraphicspath}
\end{figure}

\subsection{CSL at half filling}
\subsubsection{Details for entanglement spectrum}
In the main text, we demonstrated that the ground-state energies obtained from iPESS and iDMRG converge to very similar values. Here, we further confirm the consistency between the two methods by comparing their ES.
In the cylinder iDMRG simulations, we employ the grand-canonical ensemble together with $\rm{U}(1)\times\rm{SU}(2)$ symmetric tensors. For cylinder circumferences $L_y=4m$, antiperiodic boundary conditions are imposed along the $y$ direction, while for $L_y=4m+2$, periodic boundary conditions are used. This choice ensures that the ES of the ground state is always dominated by integer-spin IRREPs, corresponding to the identity topological sector of the CSL phase~\cite{wu2020tensor}.

Figure~\ref{fig:SM_ES_DMRG} shows the ES at $U=8$ and $U=14$ obtained using both iPESS and iDMRG. For iDMRG, we explicitly plot the dominant charge sector $Q=0$ and the subleading sectors $Q=\pm2$, indicated by black solid and green dashed lines, respectively. In contrast, the iPESS calculations retain only $\rm{SU}(2)$ symmetry, with the charge quantum number implicitly encoded through the grand-canonical formulation. Despite this difference, the level degeneracies and branch dispersions of the ES obtained from iPESS agree remarkably well with those from iDMRG, demonstrating the consistency and reliability of both methods.

We now explain the momentum location of the $Q=\pm2$ subleading branches in the Chern insulator phase in the weakly interacting limit, where the onsite interaction can be neglected. On a cylinder with antiperiodic boundary condition, the transverse momentum is quantized as $k_y/2\pi = (n + 1/2)/L_y$. In the thermodynamic limit $L_y \to \infty$, the model hosts an exact zero-energy edge mode at $k_y/2\pi = 1/4$. For finite $L_y$, this mode splits into a highest occupied negative-energy mode at $k_y/2\pi = 1/4 - 1/(2L_y)$ and a lowest unoccupied positive-energy mode at $k_y/2\pi = 1/4 + 1/(2L_y)$ (assuming positive chirality).
The ground state is obtained by filling all negative-energy modes up to $k_y/2\pi = 1/4 - 1/(2L_y)$, while leaving the positive-energy mode empty. The $Q=\pm2$ excited states correspond to adding or removing two particles at the modes closest to zero energy, with both spin components occupied (or emptied) since interaction is neglected. This leads to a total momentum shift
\[
\Delta k_y/2\pi = \pm 2 \times \left( \frac{1}{4} \pm \frac{1}{2L_y} \right),
\]
where the factor of $2$ accounts for spin degeneracy. Consequently, the $Q=\pm2$ branches appear at $k_y/2\pi = 5/8$.

\subsubsection{Robustness of CSL}

\begin{figure}[h]
\begin{localgraphicspath}{{figs/draft_fig/}}
\includegraphics[width=0.5\columnwidth]{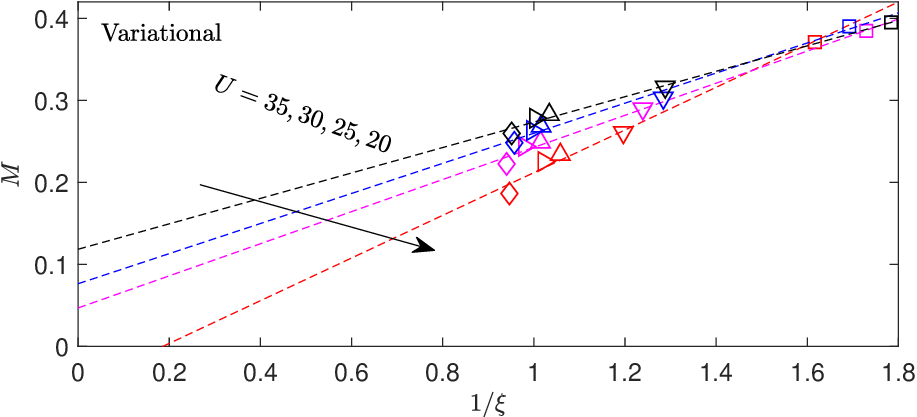}
\caption{{\bf Finite-correlation-length scaling $Z_2$ symmetric $6\times 6$ iPESS for magnetic transition.} The square, down-triangle, up-triangle, right-triangle, diamond, star symbols correspond to bond dimensions $D=6,7,8,9,10$. The linear fitting shows extrapolated magnetization versus $U$ from variational optimization.}
\label{fig:SM_mag_a}
\end{localgraphicspath}
\end{figure}

With the Mott transition point between weak-$U$ Chern insulator and intermediate-$U$ CSL fixed, here we determine the magnetic transition point between the intermediate-$U$ CSL and large-$U$ magnetic phase. In the large-$U$ limit, the Hofstadter-Hubbard model reduces to an effective NN Heisenberg model, 
favoring $120^{\circ}$ N\'eel order~\cite{gong2017global,wietek2017chiral}. 
As N\'eel order spontaneously breaks spin-rotation symmetry in 2D infinite system, 
the correlation length of spin-spin correlations in the N\'eel phase diverge. In infinite tensor network simulations, 
finite bond dimensions limits the correlation length, making finite-correlation-length scaling of 
magnetization important~\cite{rader2018finite,corboz2018finite,hasik2021investigation,ferrari2023static,hasik2024incommensurate}.
We perform $Z_2$ symmetric  iPESS simulations with a unit cell $6\times 6$ which accommodates $120^{\circ}$ N\'eel order. 
 
We analyze the data based on the empirical scaling hypothesis~\cite{hasik2024incommensurate} for $120^{\circ}$ N\'eel order
\begin{align}
M(\xi) &= M(\infty) + b/\xi,
\label{eq:scaling}
\end{align}
where $M$ scales linearly with $1/\xi$. 
Magnetization for different $U$ values appears in Fig.~\ref{fig:SM_mag_a}. The data extrapolates to negative $M$ at $U=20$, indicating a non-magnetic CSL phase. The finite-$D$ magnetization suggests proximity to a gapless critical point, similar to gapless spin liquids~\cite{liao2017gapless} where symmetry breaking can lower energy in restricted variational spaces. 
While at $U=25,30,35$, $M$ scales to a positive value at $1/\xi\rightarrow 0$ limit, which increases with $U$. The magnetic critical point is identified as  $U_{c_2}\approx 22.5$ roughly consistent  with estimates 
from the effective $J-J_{\chi}$ spin model~\cite{gong2017global,wietek2017chiral,kuhlenkamp2024chiral,divic2025anyon}. These results confirm the CSL as a robust phase for a wide parameter regime in thermodynamic limit in the Hofstadter-Hubbard model.

\subsection{Robustness of chiral SC at finite doping against competing orders}
At finite doping, charge density wave (CDW) and magnetic orders are natural competing instabilities to superconductivity. To assess the robustness of the chiral SC order, we relax the $\rm{SU}(2)$ spin symmetry down to the minimal fermion-parity $Z_2$ symmetry and consider unit-cells larger than $2\times2$. Starting from the half-filled solution, we perform variational optimizations at finite negative chemical potential $\mu$.

Throughout the optimization process, the states remain spatially uniform and non-magnetic, while the mean particle density decreases continuously and the chiral SC order parameter gradually develops. Representative data for $U=9$ and $\delta\simeq0.09$ (corresponding to $\mu=-2.5t$) are summarized in Table~\ref{table:SC}. The $2\times2$ $\rm{SU}(2)$-symmetric ans\"atz yields a slightly higher variational energy, reflecting its more restricted variational space. Nevertheless, the resulting doping level $\delta$ and pairing amplitude $|\Delta|$ are very close across all considered ans\"atze. Moreover, both the charge-order amplitude and magnetization remain negligible [$O(10^{-3})$], much smaller than the pairing amplitude $|\Delta|$.
These results demonstrate that the chiral SC order is robust against competing CDW and magnetic orders at low doping.

\begin{table}[h]
\begin{tabular}{|c|c|c|c|c|c|}
\hline
  ans\"atz & Energy & doping $\delta$ & Pairing $|\Delta|$ & $O_{\rm{CDW}}$ & $M$  \tabularnewline
\hline
  $2\times2$, $\rm{SU}(2)$ & $-0.5536$ & $0.086$ &$0.056$& $8.04e-4$& 0  \tabularnewline
\hline
  $2\times2$, $Z_2$ & $-0.5558 $ & $0.089 $ & $0.056$ & $2.8e-3$ & $1.5e-3$  \tabularnewline
\hline
  $4\times4$, $Z_2$ & $ -0.5554$ & $ 0.088$ & $0.056$ & $3.1e-3$ & $2.4e-3$  \tabularnewline
\hline
  $2\times6$, $Z_2$ & $ -0.5556$ & $ 0.088$ & $0.056$ & $2.8e-3$ & $2.3e-3$  \tabularnewline
\hline
  $6\times2$, $Z_2$ & $ -0.5556$ & $ 0.088$ & $0.056$ & $4.1e-3$ & $2.1e-3$  \tabularnewline
\hline
\end{tabular}\caption{\label{table:SC} {\bf Robustness of SC order simulated with different unit-cells and virtual symmetries.}  Here $(U, \mu, D)=(9, -2.5, 8)$ are chosen. The variational optimizations take the half-filled $U=9$ Chern insulator as the initial state.  Charge order is defined as $O_{\rm{CDW}}=\max_i{n_i}-\min_i{n_i}$, and magnetization $M$ is defined as average of single-site spin lengths $M_i={\sqrt{(s_i^x)^2+(s_i^y)^2+(s_i^z)^2}}$. }
\end{table}

\subsection{Finite-$D$ extrapolation of the superconducting order parameter}

In Fig.~4(a) of the main text, the superconducting pairing amplitude is extrapolated linearly in $1/D$. This linear form is an empirical
extrapolation ansatz rather than a universal finite-$D$ scaling law. A more physically motivated approach would be finite-correlation-length scaling~\cite{rader2018finite,corboz2018finite}. For the present chiral states, however, the finite-$D$ correlation functions
contain the gossamer-tail artifact associated with the chiral PEPS obstruction, and the corresponding correlation length does not provide a
sufficiently reliable scaling variable.

\begin{figure}[t]
\centering
\includegraphics[width=0.68\columnwidth]{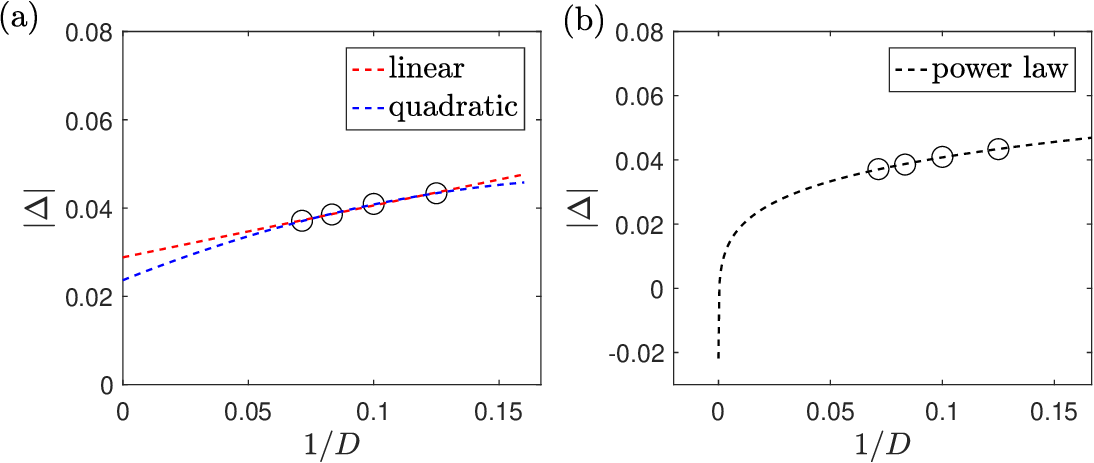}
\caption{
Finite-$D$ extrapolation of the superconducting pairing amplitude $|\Delta|$ at $U=9$ and $\delta\simeq0.05$, using the data at
$D=8,10,12,$ and $14$. The circles denote the finite-$D$ numerical results. (a) Linear fit and quadratic fit. (b) Power-law fit.
}
\label{fig:SC_extrapolation}
\end{figure}

To examine the dependence of the extrapolated pairing amplitude on the fitting ansatz, we compare the linear form
\begin{equation}
    \Delta(D)=\Delta_{\infty}+\frac{a}{D},
\end{equation}
with the quadratic form
\begin{equation}
    \Delta(D)=\Delta_{\infty}+\frac{a}{D}+\frac{b}{D^2}.
\end{equation}
As shown in Fig.~\ref{fig:SC_extrapolation}, both fits yield a finite positive $\Delta_{\infty}$, with only a modest difference between the
extrapolated values. The conclusion of a finite superconducting order parameter in the $D\rightarrow\infty$ limit is therefore insensitive to
these two fitting choices.

We also test the free-exponent power-law form
\begin{equation}
    \Delta(D)=\Delta_{\infty}+\frac{a}{D^\alpha}.
\end{equation}
With only four available bond dimensions, $D=8,10,12,$ and $14$, this three-parameter fit is poorly constrained. Although it reproduces the
finite-$D$ data, the optimal fit yields $\alpha<1$ together with a negative extrapolated intercept. We therefore do not regard the free-exponent
power-law result as a reliable extrapolation of the present limited data.

\subsection{Evolution of the pairing winding near the Mott criticality}

\begin{figure}[t]
\centering
\includegraphics[width=0.7\columnwidth]{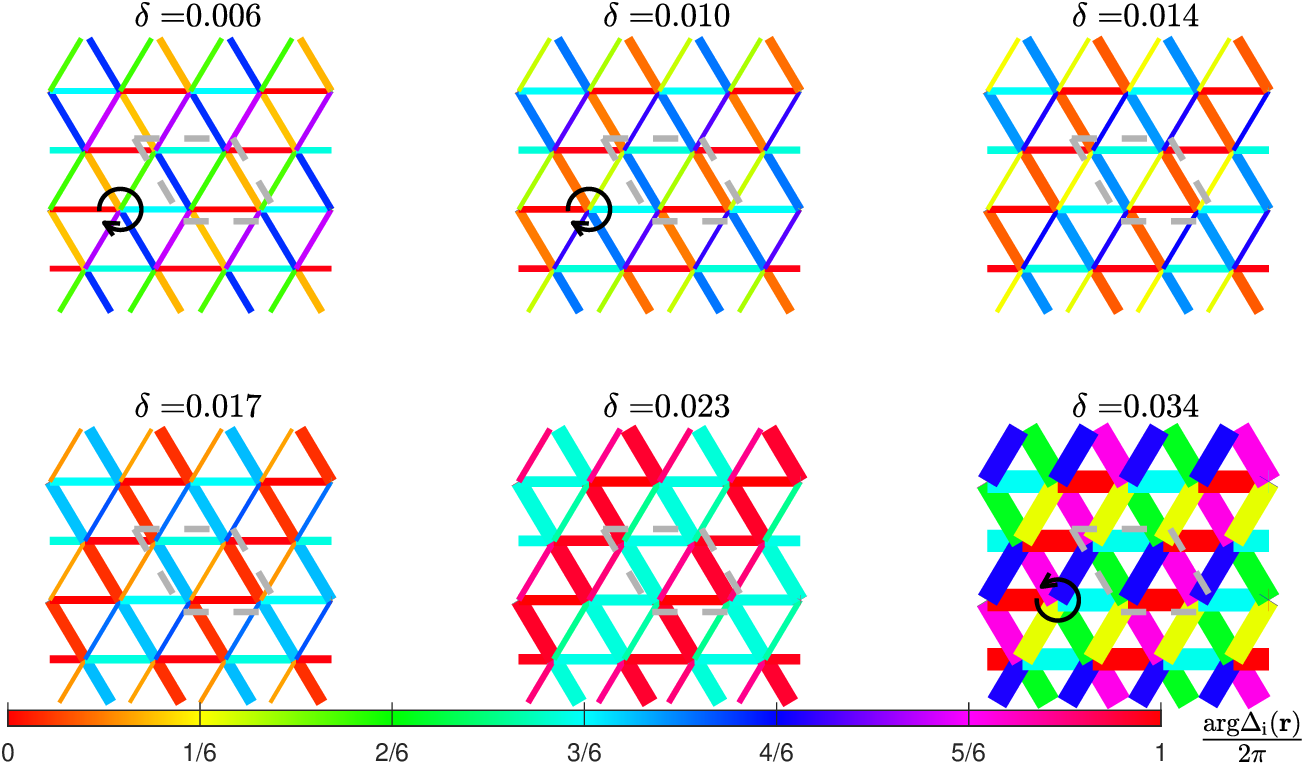}
\caption{
Evolution of the superconducting pairing pattern with hole doping $\delta$ at $U=12$ and $D=8$. All optimizations were initialized from the same
converged state at $\delta=0.006$. The thickness and color of each bond represent the pairing amplitude $|\Delta_i(\mathbf r)|$ and phase
$\arg[\Delta_i(\mathbf r)]/(2\pi)$, respectively. The black curved arrows indicate the orientation used to track the phase evolution.
}
\label{fig:pairing_pattern_evolution}
\end{figure}

To resolve the evolution between the two opposite phase-winding patterns, we perform a finer scan of the chemical potential at $U=12$.
Figure~\ref{fig:pairing_pattern_evolution} shows the resulting real-space pairing patterns for $\delta\in[0.006,0.034]$. All optimizations at different
dopings were initialized from the same converged state at $\delta=0.006$.

At the two endpoints, $\delta=0.006$ and $0.034$, the bond-resolved pairing amplitudes are relatively uniform, while the pairing phases advance
approximately uniformly along the orientation indicated by the black arrows. The two endpoint states therefore exhibit well-defined but opposite
phase-winding patterns.

At intermediate dopings, however, the pairing amplitudes become increasingly bond dependent, and the phase increments no longer follow either of the two
uniform winding patterns. In particular, around $\delta=0.023$, no unambiguous phase winding can be assigned to the pairing field. Therefore,
the winding reversal does not occur as a direct switch between two uniform winding patterns, but instead proceeds through an intermediate region that
mixes the two winding tendencies.

\section{Algorithm benchmarks}

\subsection{Symmetry qualities: iPESS vs. iPEPS }
We first demonstrate the symmetry advantage of fermionic iPESS over conventional fermionic iPEPS. As illustrated in Fig.~1(a)-(b) of the main text, in iPESS each triangle site tensor $\hat{S}_r$ connects to its nearest-neighbor (NN) site tensors $\hat{S}_{r \pm e_i}$ via an intermediate virtual tensor $\hat{T}$. This construction treats the entanglement between all NN physical sites on equal footing. In contrast, for a generic iPEPS, the bond dimension of the $M$ index after singular value decomposition is $D^2$ rather than $D$, making the entanglement along the $e_3$ edge weaker than along the other two edges.  

\begin{figure}[h]
\begin{localgraphicspath}{{figs/draft_fig/}}
\includegraphics[width=0.5\columnwidth]{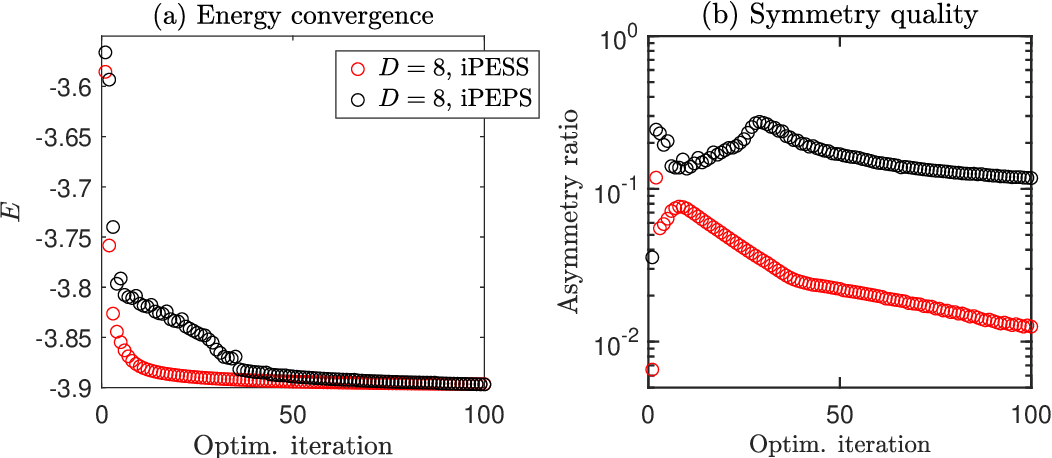}
\caption{{\bf Comparison between iPESS and iPEPS.} Here we take $(U,\mu)=(14,0)$, and the initial states is taken to be the optimized state at $(U,\mu)=(8,0)$.  (a) Convergence speed of energy during optimization. (b) Asymmetry ratio $\frac{\max{\langle c_{i}^{\dagger}c_{j}\rangle} -\min{\langle c_{i}^{\dagger}c_{j}\rangle}}{\text{mean} {\langle c_{i}^{\dagger}c_{j}\rangle}}$ of nearest neighbour hoppings.}
\label{fig:SM_compare_iPEPS_iPESS}
\end{localgraphicspath}
\end{figure}

The energy convergence and symmetry quality for both ans\"atze are shown in Fig.~\ref{fig:SM_compare_iPEPS_iPESS}. Panel (a) shows that both ans\"atze reach similar final energies, but iPESS converges faster. Although iPEPS has more variational parameters, it does not yield lower energies in practical optimization. More importantly, panel (b) shows that the variance of NN bond hopping amplitudes in iPESS is only $\sim1\%$, compared to $\sim10\%$ in iPEPS, suggesting the superior symmetry of iPESS.

\subsection{Variational method vs. simple-update method}

Next, we highlight the advantage of variational (full-update) optimization, where coarse-grained CTMRG boundary tensors are used to update bulk iPESS tensors. Gossamer correlation tails, which typically appear at a length scale $r_\text{tail} \approx 8$, are a universal feature of chiral iPEPS/iPESS. To capture chiral states, we perform $N_{\text{ite}}=50$ CTMRG iterations, effectively contracting a $2 N_{\text{ite}} m \times 2 N_{\text{ite}} n$ system, much larger than $r_\text{tail}$. This ensures an effectively infinite boundary environment, crucial for accurate variational energies. The full-update procedure (for $D>8$) remains variational because (i) the initial state is taken from prior variational optimization (up to $D=8$) rather than being random, and (ii) energy is monitored at each imaginary-time step $d\tau$, terminating when it ceases to improve.  

For comparison, we also consider the computationally cheaper simple-update method~\cite{jiang2008accurate,xie2014tensor}, where environments for tensor updates are approximated via local SVD truncations rather than coarse-grained infinite boundaries. Even with sufficiently long imaginary-time evolution, simple-update struggles to develop long-range gossamer correlations, resulting in higher energies, especially in doped regimes.

\subsubsection{Half-filling: CSL phase at intermediate-$U$}

\begin{figure}[h]
\begin{localgraphicspath}{{figs/draft_fig/}}
\includegraphics[width=1.0\columnwidth]{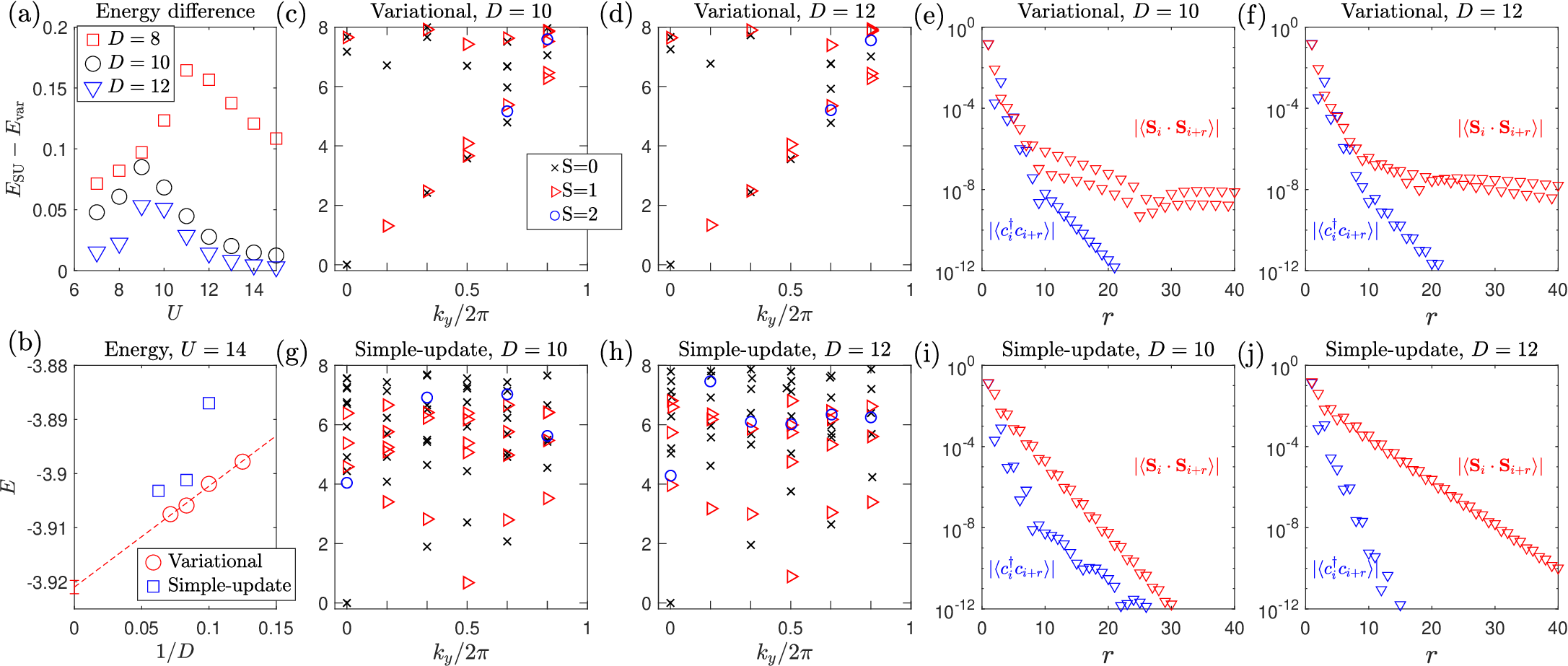}
\caption{{\bf Comparison between variational and simple-update methods at half filling.} (a) Energy difference between simple-update and variational methods. (b) $1/D$ plot of energies for $(U,\chi_{\rm{iPESS}})=(14,160)$. A heuristic linear $1/D$ fitting of the variational energy is indicated by the red dashed line. (c)-(d) and (g)-(h) Entanglement spectrum for $(U,\chi_{\rm{iPESS}})=(14,60)$. (e)-(f) and (i)-(j) Correlation functions for $(U,\chi_{\rm{iPESS}})=(14,160)$.}
\label{fig:SM_SU_CSL}
\end{localgraphicspath}
\end{figure}

Fig.~\ref{fig:SM_SU_CSL}(a) shows that simple-update energies are higher than variational energies, particularly near the Mott criticality $U_{c_1}=11.5$, indicating that simple-update does not fully exploit the variational parameter space.  

Focusing on $U=14$ in the CSL regime [Fig.~\ref{fig:SM_SU_CSL}(b)-(j)], the $1/D$ plot (panel (b)) shows that variational energies scale approximately linearly with $1/D$ and lie below simple-update energies. Panels (c)-(f) and (g)-(j) compare the ES and correlation functions from both methods. For the variational method, a minimal bond dimension $D_\text{min}=8$ is required for the optimized states to exhibit characteristic CSL ES (panels (c)-(d) for $D=10,12$). The  gossamer spin-spin correlation tails emerging for $r>r_\text{tail}$ with $r_\text{tail}\approx 8$, consistent with prior iPEPS simulations of CSLs in spin models~\cite{hasik2022simulating,niu2024chiral,niu2022chiral,chen2025simulating,weerda2024fractional}. 
In contrast to the variational method, the simple-update method does not fully capture a CSL for practical bond dimensions $D=10,12$, as shown in panels (g)-(j). The entanglement spectra remain approximately symmetric around $k_y=\pi$, although the slight enhancement of reflection-symmetry breaking at $D=12$ compared to $D=10$ indicates a trend toward chiral behavior. Similarly, the correlation functions gradually develop longer-range features, as the slope in the regime $r>r_{\rm{tail}}$ decreases with increasing $D$. These observations suggest that, at sufficiently large bond dimensions, the simple-update state would gradually develop into a CSL, analogous to the variationally optimized states shown in panels (c)-(f), although reaching such $D$ is likely beyond practical numerical limits.

\subsubsection{Half-filling: $120^{\circ}$ N\'eel phase a large-$U$}

\begin{figure}[h]
\begin{localgraphicspath}{{figs/draft_fig/}}
\includegraphics[width=0.5\columnwidth]{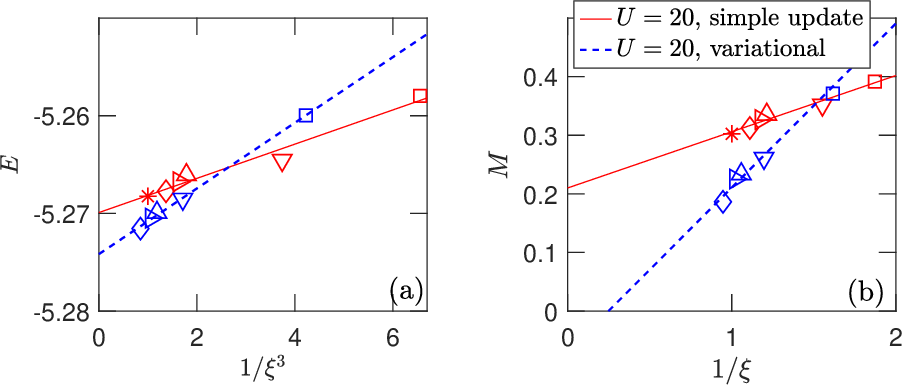}
\caption{{\bf Comparison between variational and simple-update methods for extrapolation of magnetization.} An example for finite-correlation-length scaling $Z_2$ symmetric $6\times 6$ iPESS ansatz is given for $U=20$. The square, down-triangle, up-triangle, right-triangle, diamond, star symbols correspond to bond dimensions $D=6,7,8,9,10,12$. (a)-(b) Scaling for energy and magnetization, respectively.}
\label{fig:SM_mag_b}
\end{localgraphicspath}
\end{figure}

The variational method is not only essential for simulating chiral topological states, but also important for accurate determination of magnetic transition. 
Using the empirical scaling hypothesis for $120^{\circ}$ N\'eel order
\begin{align}
E(\xi) &= E(\infty) + a/\xi^{3}, \notag \\
M(\xi) &= M(\infty) + b/\xi,
\label{eq:scaling}
\end{align}
we show the comparison of energy and magnetization scaling for $U=20$ in Fig.~\ref{fig:SM_mag_b}. 
The accurate variational data extrapolates to negative $M$ at $U=20$, indicating a non-magnetic CSL phase. In contrast, the simple update date converge to higher energy with a finite nonzero $|M|$. This highlights the importance of full variational optimization: a simple-update incorrectly yields a magnetic order and overestimated $M$.

\subsubsection{Finite doping}

\begin{figure}[h]
\begin{localgraphicspath}{{figs/draft_fig/}}
\includegraphics[width=0.7\columnwidth]{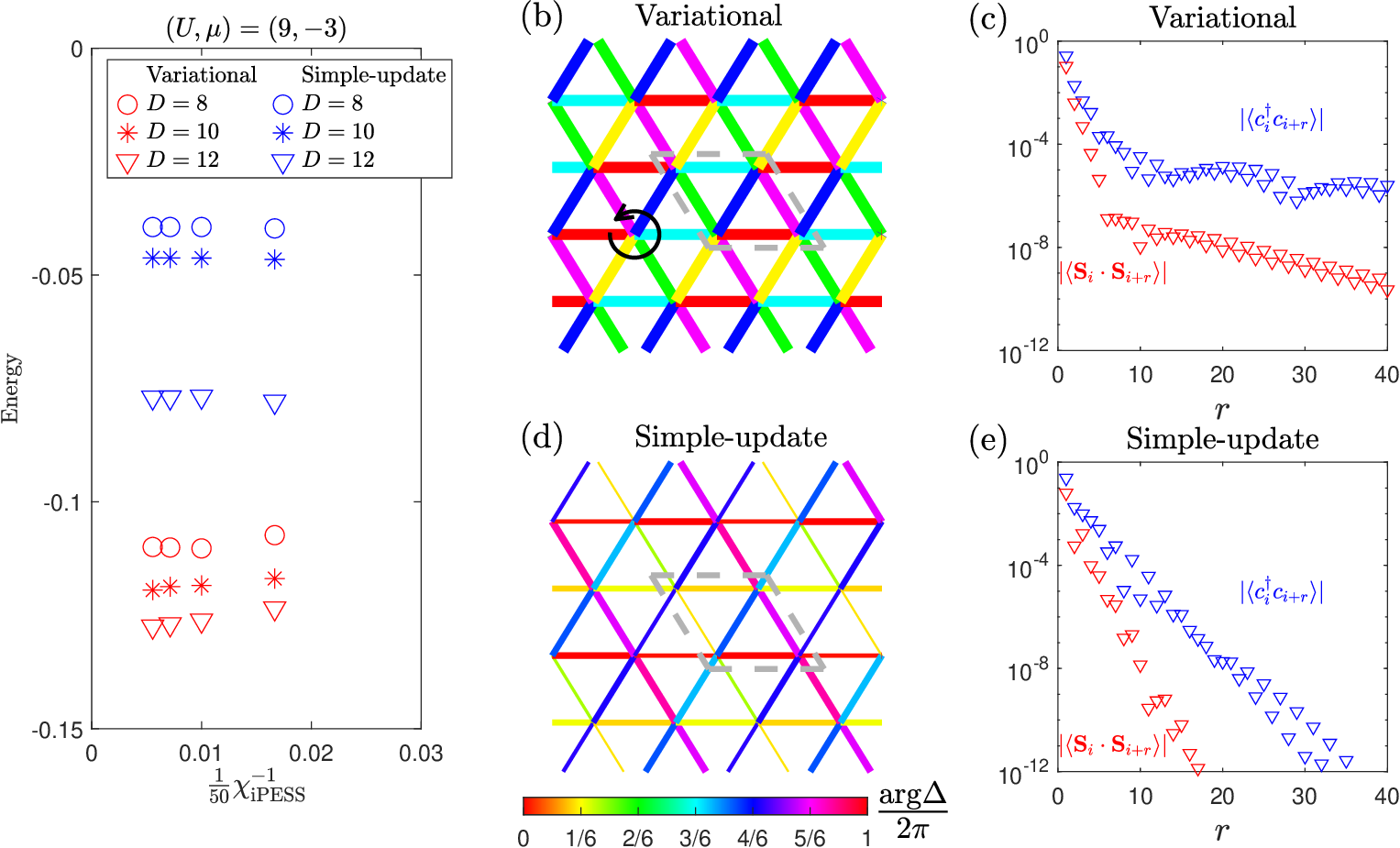}
\caption{{\bf Comparison between variational and simple-update methods at finite doping.} The parameters are chosen to be $(U,\mu)=(9,-3)$. (a) Energy difference between simple-update and variational methods. (b), (d) nearest neighbour SC pairing for $(D,\chi_{\rm{iPESS}})=(12,160)$. The doping is $\delta=0.127$ for the state obtained from variational optimization, and $\delta=0.167$ for the state obtained from simple-update method. (c), (e) Correlation functions for $(D,\chi_{\rm{iPESS}})=(12,160)$.}
\label{fig:SM_SU_dope}
\end{localgraphicspath}
\end{figure}

At finite doping, the energy difference between variational and simple-update methods becomes even more pronounced, as illustrated in Fig.~\ref{fig:SM_SU_dope}(a) for $(U,\mu)=(9,-3)$. This increase can be attributed to two factors: (i) the gapless charge degrees of freedom at finite doping induce longer-range correlations, requiring more accurate evaluation of environment tensors; (ii) the presence of an additional parameter $\bar{n}=1-\delta$ further amplifies the energy error in the simple-update method via the $-\mu \sum_\sigma n_{i\sigma}$ term in the grand canonical Hamiltonian. To exclude the possibility that simple-update results are trapped in local minima, we also initialized the iterations from the variationally optimized state, finding that the outcomes are insensitive to the initial choice.

We then examine the properties of the doped states produced by the two methods. Figures~\ref{fig:SM_SU_dope}(b)-(c) show the NN SC pairings and correlation functions for the variationally optimized states. The SC pairing amplitude is highly uniform, and its phase evolves smoothly from $0$ to $2\pi$, as indicated by the circular arrow. As in the CI and CSL phases at half-filling, the correlation functions display gossamer tails, consistent with a chiral topological SC. In contrast, the doped states obtained via simple-update [Fig.~\ref{fig:SM_SU_dope}(d)-(e)] appear spurious: the pairing amplitudes are strongly non-uniform, the phases are irregular, and the correlation functions decay essentially exponentially, reflecting the absence of a well-developed chiral SC order parameter.

\bibliography{draft.bib}

\end{document}